\documentstyle[12pt,epsfig,axodraw]{article}

\newcommand{\lsim}{\raisebox{-0.13cm}{~\shortstack{$<$ \\[-0.07cm] $\sim$}}~}
\newcommand{\gsim}{\raisebox{-0.13cm}{~\shortstack{$>$ \\[-0.07cm] $\sim$}}~}

\newcommand{\ra}{\rightarrow}
\newcommand{\ee}{e^+e^-}
\newcommand{\s}{\\ \vspace*{-3mm} }
\newcommand{\nn}{\noindent}
\newcommand{\non}{\nonumber}
\newcommand{\beq}{\begin{eqnarray}}
\newcommand{\eeq}{\end{eqnarray}}

\newcommand{\tb}{\tan\beta}

\newcommand{\ct}[1]{c_{\theta_#1}}
\newcommand{\st}[1]{s_{\theta_#1}}

\newcommand{\dt}{{\rm d}t}

\def\tth{\tilde{t}\tilde{t}h}

\def\t1{\tilde{t_1}}

\begin{document}

\def\thefootnote{\fnsymbol{footnote}}

\begin{flushright}
PM/98--40\\
\end{flushright}

\vspace{1cm}

\begin{center}

{\large\sc {\bf Associated Production of Higgs Bosons with Scalar Quarks}}

\vspace*{3mm}

{\large\sc {\bf at Future Hadron and e$^+$e$^-$ Colliders}} 

\vspace{1cm}

{\sc A. Djouadi, J.L. Kneur and G. Moultaka}

\vspace{0.5cm}

Laboratoire de Physique Math\'ematique et Th\'eorique, UMR5825--CNRS,\\
Universit\'e de Montpellier II, F--34095 Montpellier Cedex 5, France.

\end{center}

\vspace{2cm}

\begin{abstract}

\nn We analyze the production of neutral Higgs particles in association with 
the supersymmetric scalar partners of the third generation quarks at future
high--energy hadron colliders [upgraded Tevatron, LHC] and $e^+e^-$ linear 
machines [including the $\gamma \gamma$ option]. In the Minimal Supersymmetric 
extension of the Standard Model, the cross section for the associated 
production of the lightest neutral $h$ boson with the lightest top squark 
pairs can be rather substantial at high energies. This process would open a 
window for the measurement of the $h \tilde{t} \tilde{t}$ coupling, the 
potentially largest coupling in the supersymmetric theory. 

\end{abstract}

\newpage

\def\thefootnote{\arabic{footnote}}
\setcounter{footnote}{0}

\section*{1. Introduction}

One of the firm predictions of the Minimal Supersymmetric Standard Model (MSSM) 
\cite{R1} is that, among the quintet of scalar [two CP--even $h$ and $H$, a 
pseudoscalar $A$ and two charged $H^\pm$] particles contained in the extended 
Higgs sector \cite{R2}, the lightest Higgs boson $h$ should be rather light, 
with a mass $M_{h} \lsim 130$ GeV \cite{R3,R4}. This Higgs particle should
therefore be produced at the next generation of future high--energy 
colliders \cite{R6,R8}, or even before \cite{Lep2plan,tevatronplan},
if supersymmetry (at least its minimal version) is indeed realized in Nature.
At the Large Hadron Collider 
(LHC), the most promising channel \cite{R6} for detecting the $h$ boson is
the rare decay into two photons, with the Higgs particle dominantly produced  
via the gluon  fusion mechanism \cite{R7}; the two LHC collaborations 
expect to detect the narrow $\gamma \gamma$ peak for $M_h \lsim 130$ GeV, 
with an integrated luminosity $\int {\cal L}\dt \sim 300$ fb$^{-1}$ \cite{R6}. 
At future high--energy and high--luminosity $\ee$ colliders, with center of 
mass energies in the range 350--500 GeV, the properties of this particle should 
be studied in great detail \cite{R8}, to shed some light on the electroweak
symmetry breaking mechanism.  \s

In the MSSM, the genuine SUSY particles, neutralino/charginos and sfermions
could have  masses not too far from the electroweak symmetry breaking scale. In
particular the lightest neutralino, which is expected to be the lightest
SUSY particle (LSP), could have a mass in the range of $\sim 100$ GeV. Another
particle which could also be light is one of the spin--zero partners of the 
top quark, the lightest stop $\tilde{t}_1$. Indeed, because of the large
$m_t$ value, the two current stop eigenstates could strongly mix \cite{qmix}, 
leading to a mass eigenstate $\tilde{t}_1$ much lighter than the other squarks 
[which are constrained to be rather heavy \cite{data} by the negative searches 
at the Tevatron] 
and even lighter than the top quark itself. Similar features 
can also occur in the sbottom sector. These particles could therefore be also 
easily accessible at the next generation of hadron and $\ee$ colliders. \s 

If the mixing between third generation squarks is large, stops/sbottoms can 
not only be rather light but at the same time their couplings to Higgs bosons 
can be strongly enhanced. In particular, the (normalized) $h \tilde{t}_1 
\tilde{t}_1$ coupling can be the largest electroweak coupling in the MSSM. 
This might have a rather large impact on the phenomenology of the MSSM Higgs 
bosons as was stressed in Ref.~\cite{X1}.
The measurement of this important coupling would open a 
window to probe directly some of the soft--SUSY breaking terms of the 
potential. To measure Higgs--squarks couplings directly, one needs to consider 
the three--body associated production of Higgs bosons with scalar quark pairs
which has been studied recently in \cite{X2,X3,X4}. 
This is similar to the processes of Higgs boson radiation from the top quark 
lines at hadron \cite{pptth} and $\ee$ \cite{eetth} colliders, which allow to 
probe the $t\bar{t}$--Higgs Yukawa coupling directly.  \s

In this paper we extend our previous study  \cite{X2} and discuss the 
production of neutral Higgs particles in association with third generation 
scalar quarks at future high--energy 
hadron  and $e^+e^-$ linear machines [including the $\gamma \gamma$ 
option] in the unconstrained MSSM and minimal SUGRA. 
In the next section, we will summarize the properties
of third generation squarks and their couplings [including the constraints on
the latter]. In section 3 and 4, we discuss the associated production of squarks
and Higgs bosons at hadron and $\ee$ colliders, respectively. Section 5 is 
devoted to the conclusions. 

\section*{2. The Physical Set--Up} 

There are two main reasons which make the rate for the associated production 
of Higgs bosons with third generation scalar quarks potentially substantial: due
to the large Yukawa couplings, the lightest top and bottom squarks can have
relatively small masses, and their couplings to Higgs bosons can possibly be 
large. In this section, we will first summarize the properties of top and
bottom squarks and discuss their masses, mixing and their couplings to the 
MSSM neutral Higgs bosons [and to the massive gauge bosons], and then list the 
various experimental and theoretical constraints which will be taken into
account in our analysis. 

\subsection*{2.1 Masses and couplings}

In the case of the third generation, the left-- and right--handed sfermions 
$\tilde{q}_L$ and $\tilde{q}_R$ [the current eigenstates] can strongly mix 
\cite{qmix}; for a given squark $\tilde{q} = \tilde{t}, \tilde{b}$, the mass 
matrices which determine the mixing are given by
\begin{eqnarray}
 M_{\tilde{q}}^2 \ = \ 
\left[ \begin{array}{cc} m_{LL}^2 + m_q^2 & m_q \tilde{A}_q  
\\ m_q \tilde{A}_q & m_{RR}^2 + m_q^2 
\end{array} \right]
\end{eqnarray}
with, in terms of the soft SUSY--breaking scalar masses $m_{\tilde{q}_L}$ and  
$m_{\tilde{q}_R}$, the trilinear squark coupling $A_q$, the higgsino
mass parameter $\mu$ and $\tb =v_U/v_D$, the ratio of the vacuum expectation 
values of the two--Higgs doublet fields: 
\beq
m_{LL}^2= m_{\tilde{q}_L}^2 + (I_3^q - e_q s_W^2)\cos 2\beta\,M_Z^2  \ , \  
m_{RR}^2= m_{\tilde{q}_R}^2 + e_q s_W^2\,\cos 2\beta\,M_Z^2 \non \\
\tilde{A}_q = A_q-\mu(\tb)^{-2 I_3^q}  \hspace*{3cm} 
\eeq
$I_3^q$ and $e_q$ are the weak isospin and electric charge 
(in units of the electron charge) of the squark 
$\tilde{q}$, and $s_W^2=1-c_W^2 \equiv \sin^2\theta_W$. The mass matrices 
are diagonalized by $ 2 \times 2$ rotation matrices of angle $\theta_q$
\beq
 R^{\tilde{q}} &=&  \left( \begin{array}{cc}
     \ct{q} & -\st{q} \\ \st{q} & \ct{q}
  \end{array} \right)  \ \ \ \ , \ \ \ct{q} \equiv \cos \theta_q 
\ \ {\rm and} \ \ \st{q} \equiv \sin \theta_q 
\eeq
The mixing angles $\theta_q$ and the squark eigenstate masses are then given by 
\begin{eqnarray}
\label{stopmix}
\sin 2\theta_q = \frac{2 m_q \tilde{A}_q} { m_{\tilde{q}_1}^2
-m_{\tilde{q}_2}^2 } \ \ , \ \ 
\cos 2\theta_q = \frac{m_{LL}^2 -m_{RR}^2} 
{m_{\tilde{q}_1}^2 -m_{\tilde{q}_2}^2 } \hspace*{0.8cm}  \\
m_{\tilde{q}_{1,2}}^2 = m_q^2 + \frac{1}{2} \left[ 
m_{LL}^2 + m_{RR}^2 \mp \sqrt{
(m_{LL}^2 - m_{RR}^2)^2 + 4m_q^2 \tilde{A}_q^2 } \right] 
\end{eqnarray}
Due to the large value of $m_t$, the mixing is particularly strong in the stop 
sector, unless $\tilde A_t = 0$. 
This generates a large splitting between the masses of the two stop 
eigenstates, possibly leading to a lightest top squark much lighter than the 
other squarks and even lighter than the top quark. For large values of $\tb, 
\mu$ and/or $A_b$ the mixing in the sbottom sector can also be rather large, 
leading in this case to a possibly light $\tilde{b}_1$. \s

In the constrained MSSM or minimal SUGRA \cite{mSUGRA}, the soft SUSY 
breaking scalar masses, gaugino masses and trilinear couplings are universal
at the GUT scale; the left-- and right--handed sfermion masses are then given 
in terms of the gaugino mass parameter $m_{1/2}$, the universal scalar mass 
$m_0$ and $\tb$. For instance in the small $\tan \beta$ regime, 
the two soft top scalar masses [in which we include also the D--terms] at the 
low energy scale obtained from the 
one--loop  renormalization group evolution~\cite{rge}, 
are approximately given by~\cite{GUTrelations}: 
\begin{eqnarray}
m^2_{\tilde{t}_L} &=& 0.51 m_0^2 +5.29 m^2_{1/2} +0.35 M_Z^2 \cos2\beta \non \\
m^2_{\tilde{t}_R} &=& 0.01 m_0^2 +3.68 m^2_{1/2} +0.15 M_Z^2 \cos2\beta 
\label{smass}
\end{eqnarray} 
This shows that, in contrast with the first two generations, one has 
generically a sizeable splitting between $m^2_{\tilde{t}_L}$ and 
$m^2_{\tilde{t}_R}$ at the electroweak scale, due to the running of the 
(large) top Yukawa coupling [expect possibly in the marginal region where 
$m_0$ {\sl and} $m_{1/2}$ both become very small]. Thus degenerate left 
and right soft stop masses [at the electroweak scale] will be considered 
only in the unconstrained MSSM.   
Note that in this case, in most of the MSSM
parameter space, the mixing angle eq.~(\ref{stopmix}) 
is either close to zero [no mixing] or
to $-\pi/4$ [maximal mixing] for respectively, small and large values of
the off--diagonal entry $m_t \tilde{A}_t$ of the squark mass matrix eq.~(1). \s

The couplings of a neutral MSSM Higgs boson $\Phi =h,H,A$ to a pair of squarks 
$\tilde{q}_i \tilde{q}_j$ is given by
\beq
\label{couplingmatrix}
g_{\Phi \tilde{q}_i \tilde{q}_j } = \sum_{k,l=1}^2 \  
\left( R^{\tilde{q}} \right)_{ik}^{\rm T} \, C_{\Phi \tilde{q}
\tilde{q} }^{kl} \, \left( R^{\tilde{q}} \right)_{lj} 
\eeq
with the matrices $C_{\Phi \tilde{q} \tilde{q} }$ summarizing the couplings 
of the Higgs bosons to the squark current eigenstates; normalized to 
$2 (\sqrt{2} G_F)^{1/2} M_Z^2$ they are given by 
\beq
\label{h0couplings}
C_{h \tilde{q} \tilde{q}} = \left( \begin{array}{cc}
-\left(I_3^q - e_{q} s_W^2\right) \sin(\beta+\alpha) + \frac{m_q^2}{M_Z^2}
s_1^q & \frac{m_q}{2M_Z^2} (A_q s_1^q + \mu s_2^q) \\ \frac{m_q}{2M_Z^2} 
(A_q s_1^q + \mu s_2^q) & -e_{q} s_W^2 \sin(\beta+\alpha) + \frac{m_q^2}{M_Z^2} 
s_1^q  \end{array} \right) 
\eeq
\beq
\label{H0couplings}
C_{H \tilde{q} \tilde{q}} = \left( \begin{array}{cc}
\left(I_3^q - e_{q} s_W^2\right) \cos(\beta+\alpha) + \frac{m_q^2}{M_Z^2} 
r_1^q & \frac{m_q}{2M_Z^2} (A_q r_1^q + \mu r_2^q) \\ \frac{m_q}{2M_Z^2} 
(A_q r_1^q + \mu r_2^q) & e_{q} s_W^2 \cos(\beta+\alpha) + \frac{m_q^2}{M_Z^2}
r_1^q 
\end{array} \right) 
\eeq
\beq
\label{a0couplings}
C_{A \tilde{q} \tilde{q}} & = & -\frac{1}{2} \, \left( \begin{array}{cc}
0 & \frac{m_q}{M_Z^2} \left[ \mu + A_q (\tb)^{-2I_3^q} \right] \\
-\frac{m_q}{M_Z^2} \left[ \mu + A_q (\tb)^{-2I_3^q} \right] & 0 
\end{array} \right) 
\eeq
with $\alpha$ the mixing angle in the CP--even Higgs sector, and the 
coefficients $r^q_{1,2}$ and $s^q_{1,2}$ for top and bottom squarks given by
\beq
s_1^t = - r_2^t= \frac{ \cos \alpha}{\sin \beta} \ \  , \ \ 
s_2^t = r_1^t = \frac{ \sin \alpha}{\sin \beta} \ \ , \ \
s_1^b = r_2^b = -\frac{ \sin \alpha}{\cos \beta} \ \ , \ \ 
s_2^b = - r_1^b  =\frac{ \cos \alpha}{\cos \beta} 
\eeq
In the decoupling limit, $M_A \sim M_H \gg M_Z$, the mixing angle $\alpha$
reaches the limit $\beta -\pi/2$, and the expressions of the couplings become 
rather simple. One would then have for the $h\tilde{t} 
\tilde{t}$ couplings for instance
\begin{eqnarray}
g_{h \tilde{t}_1 \tilde{t}_1 } &=& \cos 2\beta \left[ 
\frac{1}{2} \cos^2 \theta_t - \frac{2}{3} s^2_W \cos 2 \theta_t \right] 
+ \frac{m_t^2}{M_Z^2} + \frac{1}{2} \sin 2\theta_t 
\frac{m_t \tilde{A}_t } {M_Z^2} \nonumber \\
g_{h \tilde{t}_2 \tilde{t}_2 } &=&  \cos 2\beta \left[ 
\frac{1}{2} \sin^2 \theta_t - \frac{2}{3} s_W^2 \cos 2
\theta_t \right] 
+ \frac{m_t^2}{M_Z^2} - \frac{1}{2} \sin 2\theta_t
\frac{m_t\tilde{A}_t } {M_Z^2} \non \\
g_{h \tilde{t}_1 \tilde{t}_2 } &=&  \cos 2\beta \sin 2 \theta_t \left[
 \frac{2}{3} s^2_W - \frac{1}{4} \right] 
+ \frac{1}{2} \cos 2\theta_t \frac{m_t\tilde{A}_t } {M_Z^2} 
\label{ghtt}
\end{eqnarray} 
and involve components which are proportional to $\tilde{A}_t = A_t -\mu/
\tan \beta$. For large values of the parameter $\tilde{A}_t$ which 
incidentally make the $\tilde{t}$ mixing angle almost maximal, $|\sin 2 
\theta_{\tilde{t}}| \simeq 1$, the last components can strongly  enhance the 
$g_{h\tilde{t}_1 \tilde{t}_1}$ coupling and make it larger than 
the top quark coupling of the $h$ boson, $g_{htt} \propto m_t/M_Z$. The 
couplings
of the heavy $H$ boson to stops involve also components which can be large;
in the case of the lightest stop, the coupling reads in the decoupling 
limit:
\begin{eqnarray}
g_{H \tilde{t}_1 \tilde{t}_1 } = \sin 2\beta \left[ 
\frac{1}{2} \cos^2 \theta_{\tilde{t}} - \frac{2}{3} s^2_W \cos 2
\theta_{\tilde{t}} \right] - \frac{m_t^2}{M_Z^2} \frac{1}{\tb}
- \frac{1}{2} \sin 2\theta_{\tilde{t}} \frac{m_t}{M_Z^2} (\frac{A_t}{\tb} 
+\mu)
\end{eqnarray}
For large $\tb$ values, the $m_t^2$ and the $A_t$ components of the
$g_{H \tilde{t}_1 \tilde{t}_1 }$ coupling are 
suppressed; only the component proportional to $\mu$ is untouched. The 
pseudoscalar $A$ boson couples only to $\tilde{t}_1 \tilde{t}_2$ pairs because 
of CP--invariance, the coupling is given by:
\begin{eqnarray}
g_{A \tilde{t}_1 \tilde{t}_2 } &=& - \frac{1}{2} \frac{m_t}{M_Z^2} 
(A_t/\tb +\mu) 
\end{eqnarray}
In the maximal mixing case, $|\sin2\theta_{\tilde{t}}| \simeq 1$, this is also 
the main component of the $H$ boson coupling to $\tilde{t}_1 \tilde{t}_2$ 
pairs, except that the sign of $\mu$ is reversed. \s

Finally, we will need in our analysis the squark couplings to the $Z$ boson. 
In the presence of mixing and normalized to the electric charge, the coupling 
$\Gamma_{Z\tilde{t}_i \tilde{t}_j } = e a_{ij}$, are given by
\begin{eqnarray}
a_{11}= \frac{2I_3^q c_{\theta_q}^2 - 2s_W^2 e_{\tilde{q}} }{2 c_W s_W} \ , \  
a_{22}= \frac{2I_3^q s_{\theta_q}^2 - 2s_W^2 e_{\tilde{q}} }{2 c_W s_W} \ , \
a_{12}= a_{21} = - \frac{ 2 I_3^q s_{\theta_q} c_{\theta_q} }{ 2c_W s_W}
\end{eqnarray}
The couplings to the photons, with the same normalization, are simply 
given by $e_{\tilde{q}} \delta_{ij}$. With this notation the 
vector and axial--vector
couplings of the electron to the $Z$ boson, that we will also need later, 
read [with the vertex convention $e \gamma_\mu (v_e -a_e \gamma_5) $] 
\begin{eqnarray}
v_e= \frac{-1+ 4s_W^2}{4 c_W s_W} 
\ \ & , & \ \  a_e= - \frac{1}{4 c_W s_W} 
\end{eqnarray}

\subsection*{2.2 Constraints used in the numerical analysis}

We will illustrate our numerical results in  the case of the unconstrained 
MSSM as well as in the minimal SUGRA case. We will concentrate on the
case of the lightest $h$ boson of the MSSM in the decoupling regime 
\cite{decoup}, and  
discuss only the production in association with light top squarks [in the case 
of no mixing, this corresponds to both squarks $\tilde{t}_1, \tilde{t}_2$ 
since their masses are almost equal, while in the case of large mixing this 
would correspond to the lightest top squark $\tilde{t}_1$]. That is, we 
consider the range of the MSSM input parameters in such a way that $M_h \ll 
M_H \simeq M_A \simeq M_{H^\pm}$, which also implies for the mixing angle 
$\alpha$ of the neutral CP--even Higgs sector $\alpha \simeq  \beta -\pi/2$. 
In this approximation, the behavior of the $\tth$ coupling with respect to the 
SUSY--breaking parameter $\tilde{A}_t$ is very simple, as discussed in the 
previous subsection. \s

The approximation of being close to the decoupling limit implies that we do 
not consider other Higgs production processes [although our analytical 
expressions will hold in these cases] with e.g. the heavy neutral $H$ 
or CP--odd $A$ bosons produced in association with top squarks. In any case,
these processes  
would be suppressed by phase--space well before the decoupling regime is 
reached 
[this is even more legitimate in the minimal SUGRA case, where for most 
of the parameter space allowed by present experimental limits, the MSSM Higgs 
sector turns out to be in this decoupling regime; see Ref.~\cite{decoup1} for
instance.] The production of the $h$ 
boson in association with light sbottoms [that we will also not 
consider in the numerical analysis]  
should follow the same pattern as for the stops, 
except that one has to replace the $\tilde{t}_1
\tilde{t}_1 h$ coupling by the $\tilde{b}_1 \tilde{b}_1 h$ coupling; the 
latter being strongly enhanced for large $\tb$ and $\mu$ values. \s

Note that, in general, we will use as input  the parameters $m_{\tilde{q}}, 
A_t$, $\tb$ and $\mu$. The $h$ boson mass is then calculated  as a function of 
these parameters [$M_h$ is only marginally affected by the variation of the 
parameter $\mu$] with the pseudoscalar Higgs boson mass fixed to $M_A=1$ TeV, 
and with the full radiative corrections in the improved effective  potential 
approach \cite{R4} included. \s

Let us now discuss the constraints which will be used in our numerical 
analysis. First there are of course the experimental 
bounds on the top squark and Higgs boson masses. From negative searches
at LEP2, a bound $m_{\tilde{t}_1} \gsim 80$ GeV is set on the lightest top
squark mass \cite{data}. Larger masses, up to $m_{\tilde{t}_1} \sim 120$ GeV, 
are excluded by the CDF collaboration \cite{cdfstop} 
in the case where the lightest 
neutralino is not too heavy,
$m_{\chi_1^0} \lsim 50$ GeV.\s

The general experimental lower bound on the lightest $h$ boson in the MSSM 
from LEP2 negative searches is $M_{h} \gsim 80$ GeV \cite{data}. However, in 
the decoupling limit the $h$ boson will behave as the SM Higgs particle for 
which the lower bound is larger, $M_h \gsim 90$ GeV; this will be the 
lower bound  
which will be used practically in our analysis since
we will work mostly in the decoupling regime. This calls for relatively large 
values of the parameter $\tb$ [since the maximal value of $M_h$ increases with 
increasing $\tb$] and/or large mixing in the top squark sector [the ``maximal 
mixing" scenario, in which the $h$ boson mass is maximized for a given $\tb$ 
value, is for $A_t \simeq \sqrt{6} m_{\tilde{q}}$].  \s

In this paper, we will often focus on the large $A_t$/small $m_{\tilde t_1}$ 
scenario, which gives the largest $\tilde t_1 \tilde t_1 h$ production cross
sections. However there are  theoretical as well as experimental constraints
on this scenario which should also be taken into account in the analysis: \s

$i)$ The absence of charge and color breaking minima (CCB) \cite{CCB} can put 
rather stringent bounds on the parameter $\tilde{A}_t$: for the unconstrained
MSSM case, a stringent CCB constraint for a large $A_t$ scenario 
reads~\cite{CCB}
\beq
A^2_t < 3 (m^2_{\tilde t_L} +m^2_{\tilde t_R}+ \mu^2 +m^2_{\Phi_U} )
\label{CCBcons}
\eeq
to be valid at the electroweak scale. [Here we only aim
at a qualitative treatment and do not enter into possible
improvements of condition (\ref{CCBcons}), 
see the last reference in \cite{CCB}. 
Note, however, that for the large $A_t$ values we consider, the CCB condition
from the precise comparison of the actual minima is practically equivalent to 
(\ref{CCBcons}).]  
This constraint involves
the stop sector parameters as defined in the previous subsection,
plus  $m^2_{\Phi_U}$, the mass squared term of the scalar doublet
$\Phi_U$. Since, as indicated above, we shall fix in our
analysis $\mu$ and $\tb$ (and $M_A \simeq$ 1 TeV for the decoupling
regime),  $m^2_{\Phi_U}$ is not an independent parameter and will be determined
consistently from the electroweak breaking constraints. It can be easily
seen that the constraint (\ref{CCBcons}) is in fact weakly dependent
on $\mu$ [and we also note as a general
tendency that (\ref{CCBcons}) is slightly less restrictive for 
relatively small $\tb$
values, $\tb \simeq $2--5, being however almost independent of $\tb$ for $\tb >
5$].  Clearly thus, large values
of $A_t$ call for larger values of $m_{\tilde t_L}$,
$m_{\tilde t_R}$  and therefore 
generally larger values of $m_{\tilde{t}_1}$.
For instance, in the mSUGRA case with the universality condition $ m_{\tilde
t_L} =  m_{\tilde t_R}
=m_0$ at the GUT scale,   
$A_t$ should be restricted to values smaller than $\sim 1.5 (1.2)$
TeV for $\tb \simeq 3 (30)$ and $m_{\tilde{t}_1} \simeq 200$ GeV. 
In the unconstrained MSSM case, one
may take relatively large $A_t$ values by pushing e.g.
$m_{\tilde t_R}$  to larger values, while still allowing a relatively light
$\tilde t_1$, depending on the values of $\mu$ and $\tb$.  
The price to pay, however, is that the 
$m_{\tilde t_L}$--$m_{\tilde t_R}$
splitting increases, therefore disfavoring too large a value for the $\tilde
t_1 \tilde t_1 h$ coupling, as is clear from eqs.~(\ref{stopmix}) and 
(\ref{ghtt}).  
Eq.~(\ref{CCBcons}) thus serves
as a relevant ``garde--fou"
to anticipate in which region of the parameter space
we can expect the $\tilde q \tilde q \Phi$ cross sections to be 
substantial without conflicting with the CCB constraints. 
In our numerical illustrations in the next section, we 
shall indicate more precisely whenever the CCB constraint 
(\ref{CCBcons}) becomes restrictive.
We note however that, after having delineated the most dangerous CCB
regions, quite often the CCB constraints are  
superseded either by the lower bound on the $h$ boson mass and/or
by the present indirect experimental  constraints
coming from the virtual contributions of squarks to the  electroweak
observables at the $Z$ peak. Indeed, in general  
the latter also forbid very large values
of the parameter $A_t$, as we shall discuss in more detail now. \s

$ii)$ Large values of $\tilde{A}_t$ lead to a large splitting of the top squark
masses; this breaks the custodial SU(2) symmetry, generating potentially large 
contributions to electroweak high--precision observables, and in particular to 
the $\rho$ parameter \cite{drho}, whose leading contributions are 
proportional to different combinations of
the squared squark masses, and which is
severely constrained  by LEP1 data \cite{LEPrho}. Since this is potentially
the most restrictive  constraint on a large $A_t$ scenario, we have thus
evaluated precisely  the contributions of the $(\tilde{t}, \tilde{b})$
isodoublet to $\Delta\rho \equiv \rho -1$ 
for typical choices of the parameters
corresponding to our  illustrative examples [e.g. to Figs.~2 and 3 of next
section].  Explicitly~\cite{drho}: 
\beq
 \Delta \rho (\tilde{t}, \tilde{b}) 
& = & 3 \frac{G_F}{8\pi^2\sqrt 2} \times \nonumber \\ \nonumber 
& & [c^2_t \;(c^2_b \;f(m^2_{\tilde t_1},m^2_{\tilde b_1})+
s^2_b \;f(m^2_{\tilde t_1},m^2_{\tilde b_2}) ) +        
    s^2_t \;(c^2_b \;f(m^2_{\tilde t_2},m^2_{\tilde b_1})
+s^2_b \;f(m^2_{\tilde t_2},m^2_{\tilde b_2}) )  \\  
& & - c^2_t s^2_t \;f(m^2_{\tilde t_1},m^2_{\tilde t_2})
-c^2_b s^2_b \;f(m^2_{\tilde b_1},m^2_{\tilde b_2}) ]  
\label{rhostop}
\eeq
where  $f(x,y) = x+y-2 x y/(x-y) \ln (x/y) $ and $s_i$, $c_i$ designate
the sine and cosine of the mixing angles in the stop and sbottom sectors. 
As expected, these contributions can become unacceptably large for very large
$\tilde A_t$ and small $m_{\tilde t_1}$, 
but stay however below the acceptable level, 
$\Delta \rho |_{\tilde{t},\tilde{b}} \lsim 3 \cdot 10^{-3}$ 
[which approximately 
corresponds to a 2$\sigma$ deviation from the SM expectation 
~\cite{LEPrho} ] for values of $\tilde{A}_t$ as large as 
1.4--1.5 TeV and stop masses as small as $m_{\tilde t_1} \gsim $ 130--140 GeV
[those precise limit values depend also on the values of $\tb$ and $\mu$].
We refrained however from making a systematic analysis of these low--energy
constraints, since these effects are also quite dependent on the
other parameters and in our numerical analysis 
we shall simply indicate whenever the contributions to the $\rho$ parameter
are becoming restrictive. 
We note, however, that the
magnitude of $\Delta\rho(\tilde{t}, \tilde{b}) $ in eq.~(\ref{rhostop})
is essentially driven
by the relative values of $m_t \tilde A_t$ and $m_{Q_L}$: 
$\Delta\rho(\tilde{t}, \tilde{b}) $ can be large 
when $m_t \tilde A_t \simeq m^2_{Q_L}$ while
there is a decoupling, $\Delta\rho(\tilde{t}, \tilde{b}) 
\simeq 1/m^2_{Q_L}$, 
whenever $m^2_{Q_L} \gg m_t \tilde A_t$. 
(This decoupling is expected, since for very large $m_{Q_L}$ 
there is
no more difference between the $W_\pm$ and $W_3$ one-loop self-energy
contributions). Incidentally, our parameter choice
when illustrating the unconstrained MSSM case below
tends to maximize the $\rho$
contributions (due to the simplifying assumption $m_{Q_L} \simeq m_{t_R}$,  
corresponding to $m_t \tilde A_t 
\simeq m^2_{Q_L}$), but this is not a generic situation 
in the unconstrained
MSSM, and can therefore be considered as conservative concerning the
limitation of present $\rho$-parameter constraints on such large $A_t$
scenarii.
Indeed, it is interesting to note that 
in the minimal SUGRA case, additional 
constraints on the soft--SUSY breaking scalar masses and trilinear couplings   
quite generically imply small $\Delta \rho(\tilde{t}, \tilde{b})$ values. 
For the typical set of
mSUGRA parameters chosen in the next sections, the corresponding contributions
to the $\rho$ parameter are  well below the present exclusion limit, often by
an order of magnitude. 
This can be traced back to the renormalization group (RG)
evolution~\cite{rge} starting 
with mSUGRA universality conditions at the GUT scale,
which indicates 
that in most cases $m^2_{Q_L} \gg m_t \tilde A_t$
at the electroweak scale, even when $\tilde A_t$ is relatively 
large~\footnote{This is actually correlated to the fact that 
RG evolution indicates that $m_{Q_L} \gsim m_{t_R}$ 
(or even $m_{Q_L} \gg m_{t_R}$) in most
mSUGRA cases,
and also that $\tilde A_t$ itself cannot be very large, as we 
shall discuss in more details in section 3.2.2 below.}
so that we are in the decoupling regime for 
$\Delta\rho(\tilde{t}, \tilde{b})$. 
\section*{3. Associated production at proton  colliders} 
In this section, we examine the associated production of Higgs bosons with a
squark pair at the LHC. This is an extension of the analysis presented in 
Ref.~\cite{X2} for the associated $h$ boson production with stop pairs, 
with a more detailed discussion of the calculation and physics results. In 
particular we shall illustrate both the case of the unconstrained MSSM as well 
as what is expected in the more constrained minimal SUGRA  case.

\subsection*{3.1 Analytical results}

At lowest order, i.e. at ${\cal O}( G_F \alpha_s^2)$, the final state 
$\tilde{q}_i \tilde{q}_i \Phi$ where $\Phi$ is a CP--even Higgs boson in 
$pp$ (or $p\bar{p}$) collisions is initiated by the Feynman diagrams 
shown in Fig.~1.

\bigskip

%%%%%%%%%%%%%%%%%%%%%%%%% Figure for pp collisions %%%%%%%%%%%%%%%%%%%%
{\bf 
\begin{center}
\begin{picture}(430,100)(0,0)
%1st diagram 
\Gluon(0,85)(50,85){4}{6}
\Text(15,75)[]{$g$}
\Gluon(0,15)(50,15){4}{6}
\Text(15,25)[]{$g$}
\DashLine(50,85)(50,15){5}
\DashLine(50,85)(100,85){5}
\Text(100,80)[]{$\tilde{q}_i$}
\DashLine(50,50)(100,50){5}
\Text(100,20)[]{$\tilde{q}_i$}
\DashLine(50,15)(100,15){4}
\Text(70,60)[]{$\Phi$}
\Text(120,50)[]{$+$}
% 2nd Diagram
\Gluon(150,85)(200,85){4}{6}
\Text(165,75)[]{$g$}
\Gluon(150,15)(200,15){4}{6}
\Text(165,25)[]{$g$}
\DashLine(200,85)(200,15){5}
\DashLine(200,85)(250,85){5}
\Text(250,80)[]{$\tilde{q}_i$}
\DashLine(215,15)(250,50){5}
\Text(250,20)[]{$\tilde{q}_i$}
\DashLine(200,15)(250,15){4}
\Text(240,60)[]{$\Phi$}
\Text(270,50)[]{$+$}
%3d diagram 
\Gluon(300,85)(350,85){4}{6}
\Text(315,75)[]{$g$}
\Gluon(300,15)(350,15){4}{6}
\Text(315,25)[]{$g$}
\DashLine(350,85)(400,85){5}
\DashLine(365,85)(400,50){5}
\Text(400,80)[]{$\tilde{q}_i$}
\DashLine(350,15)(350,85){5}
\Text(400,20)[]{$\tilde{q}_i$}
\DashLine(350,15)(400,15){4}
\Text(380,60)[]{$\Phi$}
\Text(424,50)[]{$+{\rm cros.}$}
\end{picture}
\end{center}
%%%%%%%%%%%%%%%%%%%%%%%%%%%%%%%%%%%%%%%%%%%%%%%%%%%
\begin{center}
\begin{picture}(430,100)(0,0)
%1st diagram 
\Text(0,50)[]{$+$}
\Gluon(0,75)(35,50){4}{6}
\Text(15,85)[]{$g$}
\Gluon(35,50)(0,25){4}{6}
\Text(15,15)[]{$g$}
\Gluon(35,50)(65,50){4}{6}
\DashLine(65,50)(100,75){5}
\Text(85,80)[]{$\tilde{q}_i$}
\DashLine(65,50)(100,25){5}
\Text(85,25)[]{$\tilde{q}_i$}
\DashLine(75,58)(100,40){4}
\Text(100,53)[]{$\Phi$}
% 2nd Diagram
\Text(120,50)[]{$+$}
\Gluon(130,75)(165,50){4}{6}
\Text(145,85)[]{$g$}
\Gluon(165,50)(130,25){4}{6}
\Text(145,15)[]{$g$}
\Gluon(165,50)(195,50){4}{6}
\DashLine(195,50)(230,75){5}
\Text(215,80)[]{$\tilde{q}_i$}
\DashLine(195,50)(240,25){5}
\Text(215,25)[]{$\tilde{q}_i$}
\DashLine(205,42)(230,60){4}
\Text(230,53)[]{$\Phi$}
\Text(245,50)[]{$+$}
%3d diagram 
\Gluon(250,75)(285,50){4}{6}
\Text(265,85)[]{$g$}
\Gluon(285,50)(250,25){4}{6}
\Text(265,15)[]{$g$}
\DashLine(285,50)(320,75){5}
\Text(305,80)[]{$\tilde{q}_i$}
\DashLine(285,50)(320,25){5}
\Text(305,25)[]{$\tilde{q}_i$}
\DashLine(295,58)(320,40){4}
\Text(320,53)[]{$\Phi$}
% 4th Diagram
\Text(335,50)[]{$+$}
\Gluon(350,75)(385,50){4}{6}
\Text(365,85)[]{$g$}
\Gluon(385,50)(350,25){4}{6}
\Text(365,15)[]{$g$}
\DashLine(385,50)(420,75){5}
\Text(405,80)[]{$\tilde{q}_i$}
\DashLine(385,50)(420,25){5}
\Text(405,25)[]{$\tilde{q}_i$}
\DashLine(395,42)(420,60){4}
\Text(420,53)[]{$\Phi$}
%\Text(440,50)[]{$+$}
\end{picture}
\end{center}
%%%%%%%%%
\begin{center}
\begin{picture}(250,100)(0,0)
%1st diagram 
\ArrowLine(0,75)(35,50)
\Text(15,85)[]{$q$}
\ArrowLine(35,50)(0,25)
\Text(15,15)[]{$\bar{q}$}
\Gluon(35,50)(85,50){4}{6}
\DashLine(85,50)(120,75){5}
\Text(105,80)[]{$\tilde{q}_i$}
\DashLine(85,50)(120,25){5}
\Text(105,25)[]{$\tilde{q}_i$}
\DashLine(95,58)(120,40){4}
\Text(120,53)[]{$\Phi$}
% 2nd Diagram
\Text(140,50)[]{$+$}
\ArrowLine(150,75)(185,50)
\Text(165,85)[]{$q$}
\ArrowLine(185,50)(150,25)
\Text(165,15)[]{$\bar{q}$}
\Gluon(185,50)(235,50){4}{6}
\DashLine(235,50)(270,75){5}
\Text(255,80)[]{$\tilde{q}_i$}
\DashLine(235,50)(270,25){5}
\Text(255,25)[]{$\tilde{q}_i$}
\DashLine(245,42)(270,60){4}
\Text(273,53)[]{$\Phi$}
\end{picture}
\end{center}
}
%\bigskip
\begin{figure}[htb]
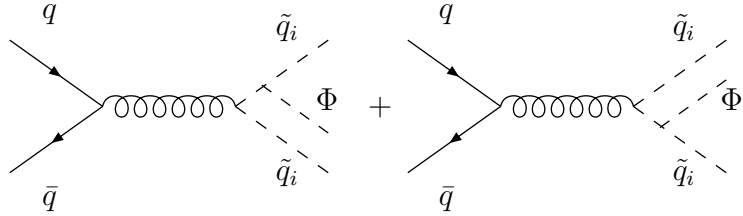

\caption[]{Feynman diagrams for the production of a CP--even Higgs
boson $\Phi$ in association with a pair of squarks via $gg$ fusion and $q
\bar{q}$ annihilation.}
\end{figure}

There are 10 diagrams  for the $gg$ fusion mechanism [including those with the 
quartic gluon--squark interaction and the three--gluon vertex] once the various
possibilities for emitting the Higgs boson from the squark lines and the  
crossing of the two gluons are added, and 2 diagrams for the $q\bar{q}$  
annihilation process\footnote{Actually, in a standard $R_\chi$ type gauge, it 
is necessary to also add two graphs, corresponding to the diagrams of Fig.~1
which involve the triple gluon vertex, but with the initial gluon lines 
replaced by ghosts. Alternatively, one can perform the calculation without 
ghosts, by choosing e.g. the axial gauge; as a cross--check of our calculation 
we have done it in both gauges and found complete agreement.}. The ${\cal O}
(G_F^3)
$ contribution from $\gamma$ and  $Z$--boson exchange diagrams, as well as the 
flavor changing contribution with a  gluino exchange diagram are negligible 
since they are suppressed, respectively, by the additional weak coupling factor
and the very small mixing between light quarks and top/bottom squarks through 
the gluino interaction, the mixing  being due to weak interactions again. \s

Due to CP--invariance which forbids the $\tilde{q}_i \tilde{q}_i A$ couplings, 
the pseudoscalar $A$ boson can only be produced in association with $\tilde{q}_
1 \tilde{q}_2$ pairs; this process [which therefore should be suppressed by 
phase space, relative to the production in association with $\tilde{q}_1 
\tilde{q}_1$ pairs] has been discussed recently in Ref.~\cite{X4}. \s

Due to the larger gluon luminosity at  high energies, 
resulting from the small values
of the parton momentum fraction $x$ that can be reached and
the large probability density of the gluon at small $x$, the contribution of 
the 
$gg$--fusion diagrams is much larger  than the contribution of the $q\bar{q}$ 
annihilation diagrams at LHC  energies: at least for most of the squark and
Higgs boson masses considered in our analysis, it  represents roughly  
95\% of the total contribution. In contrast, it is just the reverse at Tevatron
energies, where the $q\bar q$ initiated process dominates and gives about 80 \%
of the total contribution. Apart form the reduced gluon luminosity at lower 
energies, this is of course also due to the fact that Tevatron
being a $p\bar p$ collider, the $q \bar q$ hard process with valence quarks 
(antiquark) is enhanced. \s

In terms of the momenta $q_1,q_2$ of the initial gluons or quarks, and the 
momenta $p_1,p_2,p_3$ of the final two squarks and Higgs boson respectively, 
the amplitude squared $\vert {\cal M} \vert^2_{
gg}$ of the gluon initiated subprocesses $gg\ra \tilde{q}_i \tilde{q}_i \Phi$
[averaged over color and spin],  
calculated with the help of Mathematica \cite{mathematica}, is given by: 
\begin{eqnarray}
|{\cal M}|^2_{gg} &= & \frac{1}{64 \times 4} \times (4\sqrt{2} \; G_F \; M_Z^4 
\; g^2_{\Phi \tilde{q}_i \tilde{q}_i} \; g_s^4) \times \frac{16}{3} \non \\
&\times & \left[c_1 + c_2 A_1 +  c_3 A_2 + c_4 A_1 A_2  +c_5 A_1^2  + c_6 A_2^2 
\right] 
\end{eqnarray}
where $g_s$ and $G_F$ are the strong and Fermi coupling constants, and the 
coefficients $c_i$ are given by: 
\begin{eqnarray}
c_1&=& \frac{1}{4 \hat{s}^2 y_1^2 y_2^2} \bigg[-(p_1.p_2 \hat{s} 
- 2 x)^2 y_1 y_2 + 4 (m_{\tilde{q}_i}^2 \hat{s} - 4y_1) (m_{\tilde{q}_i}^2 
\hat{s} -4 y_2) (x^2 - 2 y_1 y_2) \bigg]   \nonumber \\
c_2&=& \frac{(p_1.p_2  \hat{s} - 2 x)}{4  \hat{s}^2  y_1  y_2^2} 
\bigg[ 2  m_{\tilde{q}_i}^2 \hat{s} (8  p_1.q_{-} (p_2.q_1)^2 - 8  
p_1.q_1 ((p_2.q_{+})^2 - 2  y_2) + p_1.q_{+} y_2)  \nonumber \\ 
 &&+ y_2  (-46 p_1.q_{-} (p_2.q_1)^2 + 7 \hat{s}x + 46 p_1.q_1 ((p_2.q_{+})^2 
- 2  y_2) + 10 p_1. q_{+} y_2) \bigg] \nonumber \\
%%%%%%%
c_3&=& \frac{(p_1.p_2  \hat{s} - 2 x)}{4  \hat{s}^2  y_1^2  y_2} 
\bigg[ 2  m_{\tilde{q}_i}^2 \hat{s} (8  p_2.q_{-} (p_1.q_1)^2 - 8  
p_2.q_1 ((p_1.q_{+})^2 - 2  y_1) + p_2.q_{+} y_1)  \nonumber \\ 
 &&+ y_1  (46 p_2.q_{-} (p_1.q_1)^2 + 7 \hat{s}x - 46 p_2.q_1 ((p_1.q_{+})^2 
- 2  y_1) - 10 p_2. q_{+} y_1) \bigg] \nonumber \\
%%%%%%%%%%
c_4&=& \frac{(p_1.p_2 \hat{s}-2 x)^2}{y \hat{s}^2 } + \frac{7}{2} 
p_1.q_{+} (-\frac{4}{\hat{s}} + \frac{m_{\tilde{q}_i}^2}{y_1}) + 
 \frac{1}{\hat{s}^2y_2} (9 p_1.q_{-} p_2.q_{-} - \frac{7}{2} p_2.q_{+} 
\hat{s})  (m_{\tilde{q}_i}^2  \hat{s} - 4  y_2) \nonumber \\
&&+\frac{1}{\hat{s}^2 y_1 y_2} ( (y - 8 x)  (p_1.p_2 \hat{s} -2x)^2 + 
        9  m_{\tilde{q}_i}^2 p_1.q_{-} p_2.q_{-} \hat{s}  y_2) +7 \nonumber \\
c_5&=& \frac{1}{2 \hat{s}^2 y_2^2} \bigg[2  m_{\tilde{q}_i}^2  \hat{s}^2 ( 
4 (p_2.q_{+})^2 - 9  y_2) +  (92 (p_2.q_{+})^2 + 28 p_2.q_{+} \hat{s} + 
7  \hat{s}^2 - 144  y_2)  y_2^2 \non \\
&& +  m_{\tilde{q}_i}^2  \hat{s}y_2  (-46 ( p_2.q_{+})^2 
- 7 p_2. q_{+} \hat{s} + 72  y_2) \bigg] \nonumber \\
%%%%%%%%%
c_6&=& \frac{1}{2 \hat{s}^2 y_1^2} \bigg[ 2  m_{\tilde{q}_i}^2  \hat{s}^2 (
4 (p_1.q_{+})^2 - 9  y_1) +  (92 (p_1.q_{+})^2 - 28 p_1.q_{+} \hat{s} + 
7  \hat{s}^2 - 144  y_1)  y_1^2 \non \\
&& +  m_{\tilde{q}_i}^2  \hat{s}y_1  (-46 ( p_1.q_{+})^2 
+ 7 p_1. q_{+} \hat{s} + 72  y_1) \bigg] 
\end{eqnarray}
In the previous equations, with the help of $q_{\pm}=q_1 \pm q_2$, we have used 
the following abbreviations to simplify the result: 
\begin{eqnarray}
A_1 = \frac{1}{(\hat{s}+2 p_2.q_{+})} \ \ \ , \ \ \ 
A_2 = \frac{1}{(\hat{s}-2 p_1.q_{+})} \hspace*{1cm} \nonumber 
%p_i \cdot Q_{\pm} &\equiv& p_i \cdot q_1 \pm p_i  \cdot q_2 \nonumber  
\end{eqnarray}
and
\begin{eqnarray}
x = p_1.q_2  p_2.q_1 + p_1.q_1 p_2. q_2 \ \ , \ \ 
y_i= p_i.q_1  p_i.q_2 \ \ ,  \ \ 
y = p_1.q_2  p_2.  q_2 
\end{eqnarray}
In the case of the amplitude squared $\vert {\cal M} \vert^2_{q \bar{q}}$ 
of the quark initiated subprocesses $q\bar{q} \ra \tilde{q}_i \tilde{q}_i 
\Phi$ [averaged over color and spin], 
the analytical expression is much simpler then the previous 
expression for $\vert {\cal M} \vert^2_{gg}$ and is given by:
\beq 
\vert {\cal M}\vert^2_{q \bar{q}}  = \frac{1}{9 \times 4} \; (4\sqrt{2} 
G_F M_Z^4 g^2_{\Phi \tilde{q}_i \tilde{q}_i} g_s^4) \times 
2 \times \frac{4}{\hat{s}^2} (A^2_1 d_{1} +A^2_2 d_{2} +2 A_1 A_2 d_{3}) 
\eeq
where $A_1$ and $A_2$ are given previously, and in terms of $R_1 \equiv 2 p_2 
+q_+$ and $R_2 \equiv 2 p_1 -q_+$, the coefficients $d_{i}$ are given by
\beq
d_{1} = 2 (R_1 \cdot q_1) (R_1\cdot q_2) -\frac{\hat{s}}{2} R^2_1 \ \ , \ \  
d_{2} =  2 (R_2 \cdot q_1) (R_2 \cdot q_2) -\frac{\hat{s}}{2} R^2_2 \ , \non \\
d_{3} =   (R_1 \cdot q_1) (R_2\cdot q_2) +(R_2 \cdot q_1) (R_1\cdot q_2)
 -\frac{\hat{s}}{2} (R_1\cdot R_2)  \hspace*{1cm}
\eeq	
 
The hard scattering process is convoluted with the initial parton densities, 
and integrated over the final state phase space. The resulting  
$\tilde{q}_i \tilde{q}_i \Phi$ cross--section  reads 
\begin{eqnarray}
 \sigma  (pp \to \tilde{q}_i \tilde{q}_i \Phi)  \ = \ 
 \frac{1}{2 s} \int^1_{\tau_{\rm min}} \frac{d \tau }{\tau}  
\int^1_{\tau} \frac{d x}{x} \,
 \int d {\rm PS} (q_1+\!q_2;p_1,\!p_2,\!p_3) \times 
    \nonumber \\         
\left\{ \sum_{q}
\Big[
 q (x, Q_f) 
 \bar{q} (\tau/x, Q_f) + q \leftrightarrow \bar{q}
\Big]
    \vert {\cal M} \vert^2_{q \bar{q}}    
      \,  + \,
 g (x, Q_f)   
 g (\tau/x, Q_f)  
    \vert {\cal M}\vert^2_{gg}  
\right\}                                         
\label{secpp}
\end{eqnarray}
where $q(x,Q_f)$, $\bar{q}(x,Q_f)$ [running over the five flavors $u, d, c, s, 
b$] and $g(x,Q_f)$ designate respectively the quark, antiquark and gluon 
density functions with momentum fraction $x$ and factorization scale 
$Q_f$. In eq.~(\ref{secpp}), $s$ is the hadron c.m. energy squared and as usual 
$\hat{s}= x_1 x_2\, s \equiv \tau s$ is the parton c.m. energy, with 
$\tau_{\rm min} =(2 m_{\tilde q_i} +M_\Phi)^2/s$. In eq.~(\ref{secpp}) $d$PS
$(q_1+q_2;p_1, p_2, p_3)$ is an element of the 3--body final state phase space 
normalized as 
\begin{eqnarray}
\int 
d {\rm PS} 
          &  =  &             
\int 
  (2 \pi)^4 \, \delta^4 (q_1\!+\!q_2\!-\!p_1\!-\!p_2-\!p_3) \,
  \frac{d^3 p_1}{(2\pi)^3 2 E_1} \,
  \frac{d^3 p_2}{(2\pi)^3 2 E_2} \,
  \frac{d^3 p_3}{(2\pi)^3 2 E_3}\,;         \nonumber \\
          & \to & 
  \frac{1}{(2 \pi)^4}\frac{1}{8} \, 
\int_{-1}^{+1}  d (\cos\theta)
\int_{0}^{2\pi} d \phi 
\int_{E_1,\rm min}^{E_1,\rm max} d E_1 
\int_{E_{12}, \rm min}^{E_{12}, \rm max} d E_{12} \,,
\label{phsp}
\end{eqnarray}
where $E_1$, $E_2$ are the energies of the two 
produced squarks with $E_{12} \equiv 
(E_1-E_2)/2$, and $E_3$ the energy of the Higgs particle $\Phi$. The remaining 
integral in eq.~(\ref{phsp}) is over appropriately defined angles $\theta$ and 
$\phi$, describing the motion with respect to the beam axis of the 3--momenta
{\bf p}$_1$, {\bf p}$_2$ of the two produced squarks\footnote{Actually four 
angles would 
be needed, but one angular integration is eliminated from energy conservation 
and another (azimuthal) angle integration simply gives a factor of $2\pi$ 
included in eq.~(\ref{phsp}).}. The integration bounds are
given from kinematics as follows:
\begin{equation}
  E_{1, \rm min} = m_{\tilde{q}_i} \ \ \ , \ \ \  \ \ 
  E_{1,\rm max} = \frac{\hat s + m_{\tilde{q}_i}^2 - \left(m_{\tilde{q}_i} 
+M_\Phi \right)^2}{2 \sqrt{\hat s}} 
\label{E1bound}
\end{equation}
whereas the expressions for $ E_{12,\rm min}$ and $ E_{12, \rm max} $ are more 
involved:
\begin{equation}
  E_{12, \rm min/max} = \frac{-b \mp \sqrt{b^2 - 4 a c }}{2 a }
\label{E2bound}
\end{equation}
with $a$, $b$, and $c$ given by:
\begin{eqnarray}
  a & = &  2 E_1 \sqrt{\hat s} - {\hat s} - m_{\tilde{q}_i}^2  \ \ \ , \ \ \   
  b = - \left(\sqrt{\hat s} - E_1 \right) \left(M_\Phi^2 - 
m_{\tilde{q}_i}^2 \right)
 \nonumber \\
  c & = & \frac{1}{4} \left\{ \left(E_1^2 - m_{\tilde{q}_i}^2\right)
            \left[ \left( \sqrt{\hat s} 
-E_1\right)^2 - 4 m_{\tilde{q}_i}^2 \right]
          - \left(E_1^2 + M_\Phi^2 -2 m_{\tilde{q}_i}^2  \right)^2 \right\}
\label{abc}
\end{eqnarray}
Finally, the resulting six--dimensional integral eq.~(\ref{secpp}) over the
remaining phase space and over the parton luminosities is performed numerically
with the standard Vegas Monte-Carlo integration routine~\cite{vegas}. 

\subsection*{3.2 Numerical illustrations}

We now turn to the numerical results, that we will use to illustrate the 
production of the lightest $h$ boson in the decoupling limit [with 
$M_A \simeq 1$ 
TeV]
 in association with the lightest top squarks.  
Throughout the analysis we 
will use the most recent CTEQ4 parameterizations of the parton
density functions 
\cite{CTEQ}, at leading order, with a factorization scale $Q_f = 
\sqrt {\hat s}$. Note that
we also use a (one--loop)  running $\alpha_s$, defined at the scale $\hat{s}$.
The top quark mass is fixed to $m_t=175$ GeV. 

\subsubsection*{3.2.1 Unconstrained MSSM} 

To simplify the phenomenological analysis in the case of the unconstrained 
MSSM, we will assume the left--  and right--handed stop mass parameters to be 
equal, $m_{\tilde{t}_L}= m_{\tilde{t}_R} \equiv m_{\tilde{q}}$. For
illustration, we have chosen two somewhat extreme 
values for $\tb$: $\tb =3$ and $\tb = 30$.  
Large values of $\tb$  maximize the
$h$ boson mass and therefore
circumvent the present experimental lower bound on $M_h$ 
in the decoupling limit, $M_h \gsim 90$ GeV \cite{data},
almost independently of the 
values of the parameter $A_t$ chosen in our illustrations. 
However, large $\tb$ values tend to worsen the constraints from CCB and
the $\rho$ parameters, as has been discussed in previous subsection 2.2.
For smaller 
values of $\tb$, the $g_{h \tilde{t}_1\tilde{t}_1}$ coupling is simply obtained
by rescaling the parameter $\tilde{A}_t=A_t -\mu/\tb$. An advantage of 
low $\tb$ values is 
that one can then reach relatively large values of $A_t$ and thus 
smaller $m_{\tilde t_1}$
before 
encountering problems with the CCB constraints and/or the rho parameter. 
$M_h$ also becomes 
smaller, allowing for more phase space and thus increasing the cross section,
but eventually reaching its present lowest limit for a too small value of
$A_t$. \s

In Fig.~2, the $pp \ra \tilde{t}_1 \tilde{t}_1h$ cross section [in pb] is 
displayed as a function of the lightest $\tilde{t}$ mass for the value 
$\tb=30$ in the case of no mixing $\tilde{A}_t \simeq 0$ [$A_t=0, \mu=-
100$ GeV] and  moderate mixing [$A_t=500$ GeV and $\mu=-100$ GeV]; and for
the value $\tb=3$ in the large mixing case $\tilde A_t \simeq 1.4$ TeV
[$A_t=1.2$ TeV and $\mu=-600$ GeV]\footnote{Note that for given $\tilde{A}_t$ 
and $m_{\tilde{t}_1}$ values, the cross sections in Fig.~2 are slightly 
smaller than the corresponding ones given in Ref.~\cite{X2}, where $M_h$ 
was a little lower than the 
present value due to a different choice of the parameter $\tb$.}.
For comparison the cross section for the standard--like $pp \ra \bar{t} t h$ 
process 
[that we indeed 
recalculated independently as a cross--check of our numerical
procedure] 
is of the order of 0.5 pb for a Higgs boson mass $M_h \simeq 100$ GeV 
\cite{pptth}. The cross section behaves as follows: \s

\begin{figure}[htb]
\vspace*{-0.5cm}
\begin{center}
\mbox{
\psfig{figure=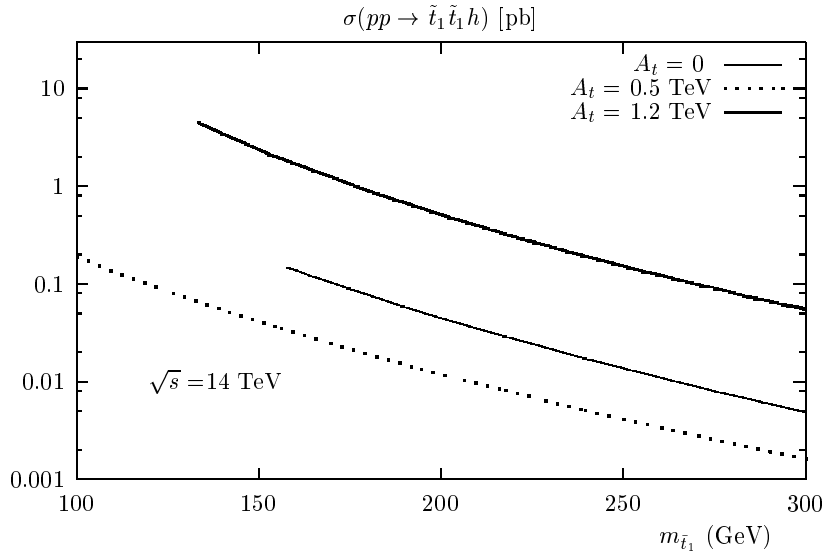,width=15cm}}
\end{center}
\vspace*{-13cm}
\caption[]{The production cross section $\sigma (pp \ra \tilde{t}_1 
\tilde{t}_1 h$) [in pb] at the LHC as a function of the $\tilde{t}_1$ mass 
and three different choices of the other 
parameters: $\tb=30$, $A_t=$ 0 (0.5) TeV;
$\tb=3$, $A_t=$ 1.2 TeV.} 
\end{figure}

-- In the case where there is no mixing in the stop sector, $\tilde{t}_1$ and
$\tilde{t}_2$ have almost the same mass [which, up to the small contribution 
of the D--terms, is constrained to be larger than $m_t^{\rm \overline{MS}}
\simeq$ 165 GeV for $m_{\tilde{q}}^2>0$] and approximately the same couplings 
to the $h$ boson since 
the $m_t^2/M_Z^2$ components in eq.~(12) are dominant. The cross section 
in Fig.~2,  which should then be multiplied by a factor of two to take  into 
account the production of both top squarks, is comparable to the cross 
section for the SM--like process $pp \ra t\bar{t} h$ in the  low mass 
range\footnote{In scenarii where the  $\tilde{t}$ masses are related to the 
masses of the light quark partners $m_{\tilde{q}}$, the mass range 
$m_{\tilde{t}}\lsim  200$ GeV in the no mixing case is, 
however, ruled out by present experimental constraints on $m_{\tilde{q}}$ 
[which approximately corresponds to the common mass of the squarks of the
first two generations and the sbottom] at 
the Tevatron \cite{data}.} $m_{\tilde{t}}\lsim  200$ GeV. \s

-- For intermediate values of $\tilde{A}_t$ the two components of the 
$h \tilde{t}_1 \tilde{t}_1$ coupling interfere destructively [with our 
conventions, $\sin 2\theta_t <0$ in the relevant parameter space,
see eq. (\ref{stopmix})] partly 
canceling each other and resulting in a small cross section, unless
$m_{\tilde{t}_1} \lsim 100$ GeV. For some value of $\tilde{A}_t$,
$g_{h\tilde{t}_1 \tilde{t}_1} \sim 0$ and the cross section vanishes. \s

-- In the large mixing case, $ A_t \sim 1.2$ TeV, $\sigma(pp\ra 
\tilde{t}_1 \tilde{t}_1 h)$ can be quite large without conflicting with
the present bounds on $M_h$, the rho parameter, or the CCB 
constraints~\footnote{The range of 
$\tilde{t}_1$ masses below $m_{\tilde{t}_1} \lsim 130$ GeV where the cross 
section is the largest, gives too large contributions to the $\rho$ parameter 
for this specific large value of $\tilde A_t$, 
and is also excluded already by the $h$ boson 
mass lower bound.}. 
It is above the rate for the 
standard process $pp \ra \bar{t}th$ for values of $m_{\tilde{t}_1}$ smaller 
than 210--220 GeV, approximately. 
If  $\tilde{t}_1$ is lighter than the top quark, the $\tilde{t}_1
\tilde{t}_1 h$ cross section significantly exceeds the one for $\bar{t}th$ 
final states. For $m_{\tilde{t}_1}=150$ GeV for instance, $\sigma( pp \ra 
\tilde{t}_1 \tilde{t}_1 h)$ is about a factor five larger than $\sigma(pp 
\ra t\bar{t}h)$. \s

\begin{figure}[htb]
\vspace*{-.8cm}
\begin{center}
\mbox{
\psfig{figure=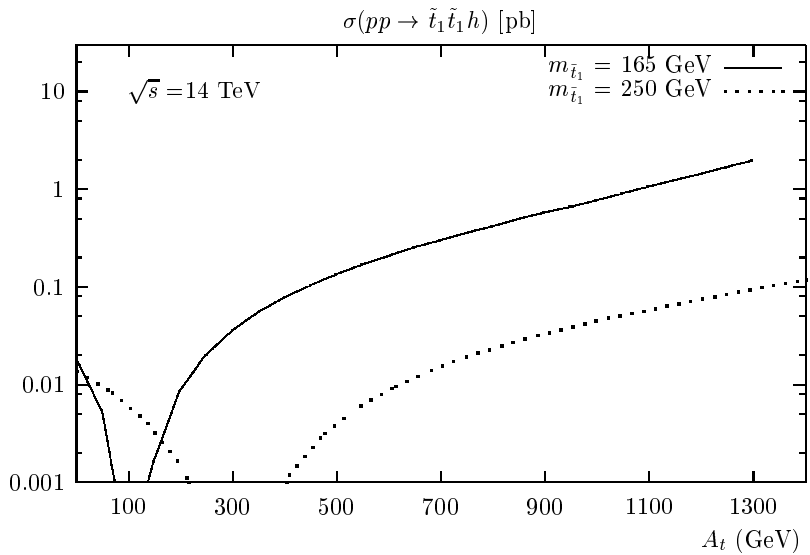,width=15cm}}
\vspace*{-13.5cm}
\end{center}
\caption[]{The cross section $\sigma(pp \ra \tilde{t}_1\tilde{t}_1h $) [in pb] 
at the LHC as a function of $A_t$ for fixed $m_{\tilde{t}_1}=$165 GeV,
$\tb=3$; 
and 250 GeV, $\tb=30$.}
\end{figure}

In Fig.~3, we fix the lightest top squark mass to $m_{\tilde{t}_1} =165$ and 
$250$ GeV,
 and display the $pp \ra gg+q\bar{q} \ra \tilde{t}_1 \tilde{t}_1h$ cross 
section as a function of $A_t$.  The value of $\tb$ and $\mu$ are fixed
to $\tb=3$, $\mu = -600$ GeV  
and $\tb=30$, $\mu =-100$ GeV respectively, 
in order to illustrate that the
$A_t$ range allowed 
from the different constraints also depends somewhat on
the values of $\tb$, as already explained above. 
For $m_{\tilde{t
}_1} =165$ GeV, the production cross section is rather small for the no mixing 
case and even smaller for the
intermediate mixing [becoming negligible for $\tilde{A}_
t$ values between 50 and 150 GeV], and then becomes very large, exceeding the 
reference cross section 
[$\sigma( pp \ra \bar{t}t h ) \sim 0.5$ pb for $M_h \sim 
100$ GeV] for values of $\tilde{A}_t$ above $\sim 0.9$ TeV. For fixed value of 
$\tilde{A}_t$, the cross section decreases with increasing $\tilde{t}_1$ mass
since the phase space becomes smaller. It is however still sizeable for
$m_{\tilde{t}_1} =250$ GeV and $\tilde{A}_t\gsim 1.4$ TeV. \s

Note that for fixed $\tilde{t}$ mass and coupling, the cross section becomes
smaller for larger values of $\tilde{A}_t$, if $\tilde{A}_t \lsim \sqrt{6} 
m_{\tilde{q}}$, because $M_h$ increases \cite{R4} and the process is less 
favored by phase--space; in the reverse situation, $\tilde{A}_t \gsim \sqrt{6} 
m_{\tilde{q}}$, the $h$ boson mass will start decreasing with increasing 
$\tilde{A}_t$ [reaching values below $M_h \lsim 90$ GeV when e.g. $A_t 
\sim 1.3$ TeV for $m_{\tilde{t}_1} =165$ GeV and $\tb=3$, where we 
thus stopped the
corresponding plot] and the phase--space 
is more favorable to the reaction.  

\subsubsection*{3.2.2 The mSUGRA case}

Let us now turn to the special case of the minimal SUGRA model~\cite{mSUGRA}, 
in which 
universality at the GUT scale implies that the only free parameters 
\cite{R1} are the 
values of $m_0$, $m_{1/2}$, $A_0$, respectively the common scalar 
mass, gaugino mass and trilinear scalar coupling, plus the sign of the $\mu$ 
parameter and $\tan\beta$. All the physical parameters, and in particular the 
squark masses, are obtained from a renormalization group (RG) 
running from the GUT 
scale to the weak scale~\cite{rge}, 
where the electroweak symmetry breaking 
constraints fix  the $|\mu|$ and $B$ parameters 
of the Higgs potential at the electroweak 
scale\footnote{The RG evolution of the relevant parameters and calculation
of the physical masses and couplings is done with 
a numerical code \cite{suspect} including in particular supersymmetric 
threshold effects at the one--loop level, and 
a consistent treatment of the  electroweak symmetry breaking.}.\s

For our numerical illustrations, Fig.~4, we choose specific values of the 
above parameters in such a way that one of the top squarks can be sufficiently 
light in order to have sizeable cross sections. In general, the cross 
section in the mSUGRA scenario may be as substantial as in the unconstrained 
MSSM cases illustrated above, 
but this occurs in a more restricted region of the parameter space. 
This is essentially due to the fact that it is generically very difficult to 
have almost degenerate $\tilde{t}_L$ and $\tilde{t}_R$ in mSUGRA as 
already discussed in section 2 and as is clear 
from eqs.~(\ref{smass}), so that the stop mixing angle which is controlled 
by the ratio $\tilde A_t/(\tilde m^2_{\tilde{t}_L} -\tilde m^2_{\tilde{t}_R})$
can become large only for very large $\tilde A_t$.
Moreover considering the 
RG equation driving the $A_t$ parameter \cite{rge}:
\beq
\frac{d A_t}{dt} =\beta(A_t) = 
\frac{1}{8\pi^2} (\sum_{i=1}^{3} c_i g^2_i M_i +6 Y^2_t A_t +
Y^2_b A_b)
\label{rgat}
\eeq
where $t \equiv \ln Q$, 
$g_i$, $Y_i$ are the gauge and  Yukawa couplings and $M_i$ the gaugino
masses [with the constants $c_i$ given by: $c_1=26/15, c_2=6$ and $c_3=32/3$],
one can see that for most  cases $\vert A_t\vert$ tends to decrease 
when the energy scale $Q$ is decreasing from 
GUT to low-energy. Indeed, the $c_i$'s in eq.~(\ref{rgat}) are all positive 
and in mSUGRA $M_i({\rm GUT}) \equiv m_{1/2} >0 $, so that $\beta(A_t) >0$ for 
$A_t >0$. This accordingly makes a large $A_t$ value at 
low energy less likely, 
since $A_0 =A_t({\rm GUT})$ would have to be even larger, which may conflict 
with the CCB constraints among other things, 
as it was discussed in the previous
section. The only way to have an increasing $\vert A_t\vert $ when 
running down to low energy is if $A_0 <0 $  with $A_0$  small enough that
$\beta(A_t)$ in  eq.~(\ref{rgat}) remains positive, but the latter positiveness
requires a relatively large $m_{1/2}$ value, which in turn  implies 
from eq.~(\ref{smass}) that $m_{\tilde t_1}$ cannot be small. 
[Note that, due to the universality of the trilinear scalar coupling,
the above features are qualitatively unchanged even when the Yukawa
bottom contribution in eq.~(\ref{rgat}) is not neglected.] \s

\begin{figure}[htb]
\vspace*{-0.5cm}
\begin{center}
\mbox{
\psfig{figure=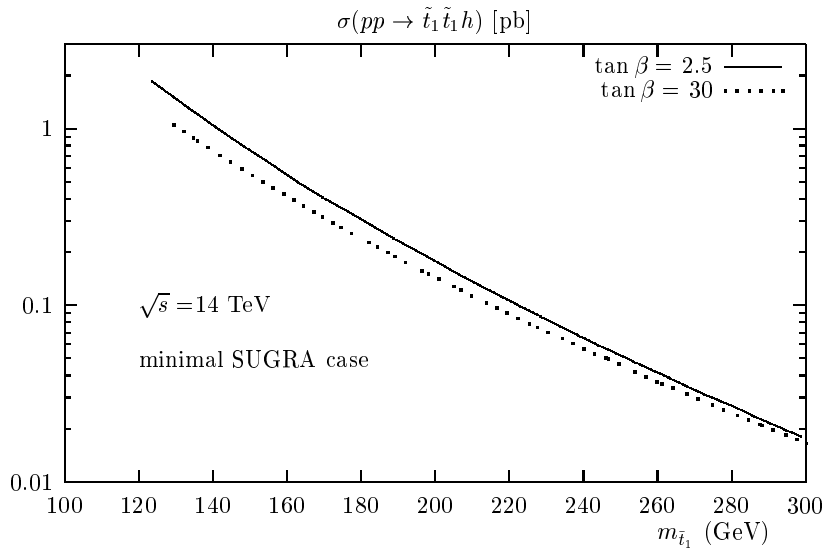,width=15cm}}
\end{center}
\vspace*{-13cm}
\caption[]{The  cross section $\sigma (pp \ra \tilde{t}_1 
\tilde{t}_1 h$) [in pb] at the LHC as a function of the $\tilde{t}_1$ 
mass in the mSUGRA case: $\tan\beta=2.5 (30)$, $m_{1/2} =$ 300 GeV, 
$A_0 = -2$ TeV, sign($\mu)=+$. }  
\end{figure}

The mSUGRA cross section as  illustrated in Fig.~4 
actually corresponds  to a situation 
where $|A_t|$ at the relevant electroweak scale 
is not very large, $|A_t| \lsim 1$ TeV, while the effective 
coupling $|\tilde A_t|$ becomes larger, at least for the small $\tan\beta 
= $2.5 case, 
because $\mu$ [which we determine consistently from radiative electroweak 
symmetry breaking] turns out to be rather large, $\mu \sim 1$ TeV. In this 
case, for a sufficiently low $m_{\tilde t_1}$ value the cross section is 
potentially as large as in the case of the unconstrained MSSM, but this 
situation clearly occurs in a restricted region of the mSUGRA parameter space.
On the other hand, as discussed in subsection 2.2,  
the mSUGRA case is accordingly less
subject to the CCB constraint, since $A_t$ is not too large, 
and also no restrictions are imposed by the  $\rho$ 
parameter until the $\tilde{t}_1$ mass becomes really small. 
However, one should take into account another constraint, 
namely that of requiring the lightest neutralino [rather than the lightest top 
squark] to be  
the lightest SUSY particle (LSP). This is the reason why in Fig.~4 the plots 
start at $m_{\tilde t_1}$ values larger than $m_{\chi_1^0} \sim 120$ GeV,
which corresponds to our choice of the basic mSUGRA parameters. 

\subsubsection*{3.2.3 Associated production at the upgraded Tevatron}

For completeness, we illustrate some values of the $\tilde t_1 \tilde t_1 h$ 
production cross section at the upgraded Tevatron with $\sqrt s =$ 2 TeV. 
Naively, one may expect that such a three particle final state process will 
be very disfavored by phase space in the parton frame, except for a very light 
$\tilde t_1$ and very light Higgs bosons which are already excluded by present
experimental constraints. However, this clear phase space suppression is 
partially compensated by the fact that the $q\bar q$ hard scattering process 
[which scales as the inverse of the parton c.m. energy $\hat{s}$] dominates 
the cross section at the Tevatron. 
More precisely, due to the fact that Tevatron is a $p\bar p$ collider, the 
cross section benefits form the  
larger quark/antiquark densities relative to the
 LHC case. \s

\begin{figure}[htbp]
\vspace*{-.5cm}
\begin{center}
\mbox{
\psfig{figure=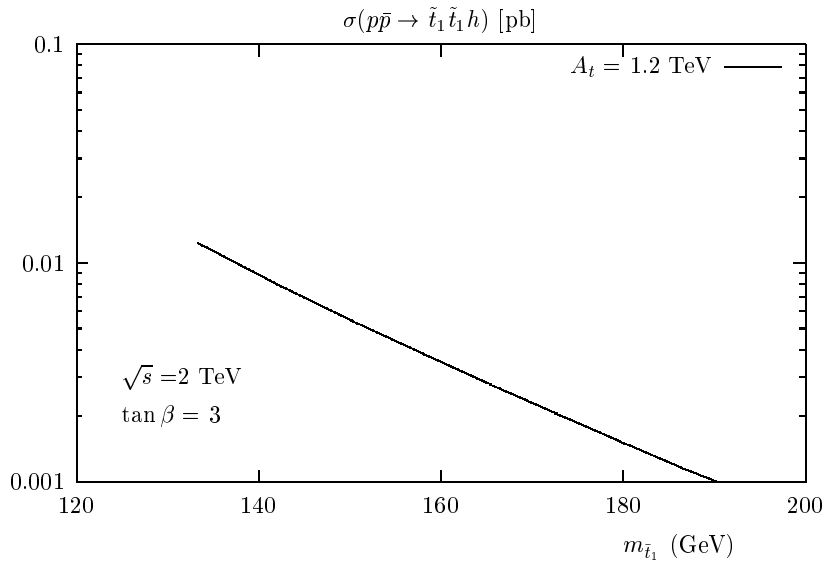,width=15cm}}
\end{center}
\vspace*{-13cm}
\caption[]{The production cross section $\sigma (p\bar p \ra \tilde{t}_1 
\tilde{t}_1 h$) at the Tevatron [in pb] 
as a function of the $\tilde{t}_1$ mass for $A_t=1.2$ TeV, 
$\mu = -$600 GeV and $\tb=3$.} 
\end{figure}

The cross section $\sigma (p\bar p \ra \tilde{t}_1 \tilde{t}_1 h$) is shown 
in Fig.~5 at a c.m. energy $\sqrt{s}=2$ TeV in the case of large $A_t$ values, 
since otherwise the cross section is clearly too small for the process
to be relevant. For the planned luminosities of $\int {\cal L}\dt \sim 10$ 
fb$^{-1}$ \cite{tevatronplan}, one can see from Fig.~5 that up to a 
few tens of events might be collected for $m_{\tilde{t}_1} \sim 130$ GeV 
[smaller masses for such a large $A_t$ value, are already ruled out
by the present constraint, as discussed above], before efficiency 
cuts are applied.  
This gives some hope that this process might be useful at the Tevatron too. 
We shall discuss in the next subsection the possible signal for this process, 
which is similar for the LHC and Tevatron. 

%%%%%%%%%%%%%%%%%%%%%%%%%%%%%%%%%%%%%%%%%%%%%%%%%%%%%%%%%%%%%%%%%%%%%%
\subsubsection*{3.2.4 Signal}

Finally, let us discuss the signal for the $pp \ra \tilde{t}_1 \tilde{t}_1 h$
process. If the mass of the lightest top squark is only slightly larger than
that of the lightest neutralino [which is expected to be the LSP], the 
dominant $\tilde{t}_1$ decay channel \cite{stopdecay} will be the loop induced 
decay into
a charm quark plus a neutralino, $\tilde{t}_1  \ra c \chi_1^0$. 
For larger $m_{\tilde{t}_1}$ values, the lightest top squark
can decay into a $b$ quark and a chargino, $\tilde{t}_1 \ra b\chi^+$
and possibly into a $t$ quark and the LSP, $\tilde{t} \ra t \chi_1^0$. 
[We assume that the strong decay into gluinos does not occur]. 
In the interesting region where  the cross section $\sigma(pp 
\ra \tilde{t}_1  \tilde{t}_1 h)$ is large, i.e. for relatively light 
$\tilde{t}_1$, the decay  mode  $\tilde{t}_1 \ra b \chi^+$ can be dominant, 
unless the mass difference  $m_{\tilde{t}_1} -m_{\chi_1^+}$ is very small.
Assuming that the  partners of the leptons are heavier than
the lightest chargino, $\chi_1^+$  will mainly decay into the LSP and a real or
virtual $W$ boson, leading to the  final state
\begin{eqnarray}
\tilde{t}_1 \ra bW^+ \ + \ {\rm missing \ energy}
\end{eqnarray}
This is the same topology as in the case of the top quark decay, $t \ra 
bW^+$, except that in the case of the top squark there is a large amount of 
missing energy due to the undetected LSP. If sleptons are also relatively 
light, charginos decay will also lead to $l \nu \chi_1^0$ final states. 
The only difference between the final states generated by the $\tilde{t}
\tilde{t}h$ and $t\bar{t}h$ processes, will be due to the softer energy 
spectrum of the charged leptons coming from the chargino decay in the former
case, because of the energy carried by the invisible neutralinos. \s

The Higgs boson can be tagged\footnote{Decays of the $h$ boson, produced in
association with $t\bar{t}$ pairs, into the main decay channel $h \ra b\bar{b}$
have also been discussed in the literature \cite{bbh}.} through its $h \ra 
\gamma \gamma$ decay mode \cite{R6}.
In the decoupling limit, and for light top squarks and large $\tilde{A}_t$
values, the branching ratio for this mode can be  enhanced
[up to $\sim 20\%$] compared to the $\gamma \gamma$ branching ratio of the
SM Higgs boson \cite{X1}, because of the additional
contributions of the $\tilde{t}$--loops which interfere constructively with 
the dominant $W$--loop contribution. \s

Although a  detailed analysis taking into account backgrounds and
detection efficiencies, which is beyond the scope of this paper, will be 
required to assess the importance of this signal, it is clear that 
in some areas of the MSSM parameter space, $\gamma \gamma$+ charged 
lepton events can be much more copious than in the SM, and the contributions 
of the $pp \ra \tilde{t} \tilde{t} h$ process to these events can render the 
detection of the $h$ boson much easier than with the process $pp \ra t 
\bar{t}h$ alone. This excess of events could then allow to measure 
the $h \tilde{t}_1 \tilde{t}_1$ coupling, and open a window to probe the
soft--SUSY breaking parameters of the stop sector. In particular one can
obtain some information on the trilinear coupling $A_t$, which is rather 
difficult to extract from the study of the production of SUSY particles at 
the LHC as  shown in Ref.~\cite{LHCscan}.

\section*{4. Associated production at e$^+$e$^-$ colliders} 

At future linear $\ee$ colliders, the final state $\tilde{q}_1 \tilde{q}_1 h$
may be generated in three ways: $(i)$ two--body production of a mixed pair
of squarks and the decay of the heaviest squark to the lightest one and a 
Higgs boson [Fig.~6a]; $(ii)$ the continuum production in $\ee$ 
annihilation\footnote{Recently, the continuum production in $\ee$ annihilation 
has also been evaluated in Ref.~\cite{X3} using the numerical graph 
calculation  program GRACE--SUSY \cite{grace}.} [Fig.~6b]; 
and $(iii)$ the continuum production in $\gamma \gamma$ collisions
[Fig.~6c]. The same picture holds for the associated production of the heavier
CP--even Higgs boson $H$ with $\tilde{q}_1 \tilde{q}_1$ pairs. In the case of 
the pseudoscalar boson $A$ only the processes $(i)$ and $(ii)$ are possible:
because of CP--invariance, the coupling $A\tilde{q}_1 \tilde{q}_1$ is absent
and the final state $\tilde{q}_1 \tilde{q}_1 A$ cannot be generated in $\gamma
\gamma$ collisions at tree--level; furthermore, in $\ee$ annihilation the 
graph where the
$A$ boson is emitted from the $Z$ line is absent, but there is another
graph involving the  $ZAh(H)$ coupling [see Fig.~6b]. For the associated 
production of CP--even Higgs bosons with $\tilde{q}_2 \tilde{q}_2$ states, 
only the processes $(ii)$ and $(iii)$ are
relevant. 
\newpage
{\bf 
\noindent {\large a)}}
\vspace*{-1.3cm}
\begin{center}
\begin{picture}(300,100)(0,0)
%1st diagram 
\ArrowLine(0,75)(35,50)
\Text(15,85)[]{$e^-$}
\ArrowLine(35,50)(0,25)
\Text(15,15)[]{$e^+$}
\Photon(35,50)(85,50){4}{6}
\Text(58,37)[]{$Z^*$}
\DashLine(85,50)(120,75){5}
\Text(105,80)[]{$\tilde{q}_1$}
\DashLine(85,50)(120,25){5}
\Text(105,25)[]{$\tilde{q}_2$}
\DashLine(185,50)(235,50){5}
\Text(205,37)[]{$\tilde{q}_2$}
\DashLine(235,50)(270,75){5}
\Text(255,80)[]{$\tilde{q}_1$}
\DashLine(235,50)(270,25){5}
\Text(255,25)[]{$\Phi$}
\end{picture}
\end{center}
%%%%%%%%%%%%%%%%%%%%%%%%%%%%%%%%%%%%%%%
\noindent {\large b)}
\vspace*{-1.3cm}
\begin{center}
\begin{picture}(400,100)(0,0)
%1st diagram 
\ArrowLine(0,75)(35,50)
\Text(15,85)[]{$e^-$}
\ArrowLine(35,50)(0,25)
\Text(15,15)[]{$e^+$}
\Photon(35,50)(85,50){4}{6}
\Text(55,37)[]{$\gamma,Z^*$}
\DashLine(85,50)(120,75){5}
\Text(105,80)[]{$\tilde{q}_i$}
\DashLine(85,50)(120,25){5}
\Text(105,25)[]{$\tilde{q}_i$}
\DashLine(95,58)(120,40){4}
\Text(120,53)[]{$\Phi$}
% 2nd Diagram
\Text(140,50)[]{$+$}
\ArrowLine(150,75)(185,50)
\Text(165,85)[]{$e^-$}
\ArrowLine(185,50)(150,25)
\Text(165,15)[]{$e^+$}
\Photon(185,50)(235,50){4}{6}
\Text(205,37)[]{$\gamma,Z^*$}
\DashLine(235,50)(270,75){5}
\Text(255,80)[]{$\tilde{q}_i$}
\DashLine(235,50)(270,25){5}
\Text(255,25)[]{$\tilde{q}_i$}
\DashLine(245,42)(270,60){4}
\Text(270,53)[]{$\Phi$}
\Text(340,50)[]{$(\Phi= h, H, A)$}
\end{picture}
\end{center}
\begin{center}
\begin{picture}(400,100)(0,0)
% 3d Diagram
\Text(0,50)[]{$+$}
\ArrowLine(0,75)(35,50)
\Text(15,85)[]{$e^-$}
\ArrowLine(35,50)(0,25)
\Text(15,15)[]{$e^+$}
\Photon(35,50)(85,50){4}{6}
\Text(60,37)[]{$Z^*$}
\Photon(85,50)(100,65){4}{3}
\DashLine(100,65)(120,85){4}
\Text(120,75)[]{$\tilde{q}_i$}
\DashLine(85,50)(120,15){5}
\Text(100,25)[]{$\Phi$}
\DashLine(100,65)(120,40){4}
\Text(120,55)[]{$\tilde{q}_i$}
% 4d Diagram
\Text(140,50)[]{$+$}
\ArrowLine(160,75)(195,50)
\Text(175,85)[]{$e^-$}
\ArrowLine(195,50)(160,25)
\Text(175,15)[]{$e^+$}
\Photon(195,50)(245,50){4}{6}
\Text(220,37)[]{$Z^*$}
\DashLine(245,50)(260,65){4}
\Text(250,65)[]{$\Phi$}
\DashLine(260,65)(280,85){4}
\Text(280,75)[]{$\tilde{q}_i$}
\DashLine(245,50)(280,15){5}
\Text(265,25)[]{$A$}
\DashLine(260,65)(280,40){4}
\Text(280,55)[]{$\tilde{q}_i$}
\Text(340,50)[]{$(\Phi= h, H)$}
\end{picture}
\end{center}
%%%%%%%%%%%%%\gamma \gamma collisions %%%%%%%%%%%%%%%%%%%%%%%%%%%%%%%%%
{\large c)}
\vspace*{-.7cm}
\begin{center}
\begin{picture}(430,100)(0,0)
\Photon(0,85)(50,85){4}{6}
\Text(15,75)[]{$\gamma$}
\Photon(0,15)(50,15){4}{6}
\Text(15,25)[]{$\gamma$}
\DashLine(50,85)(50,15){5}
\DashLine(50,85)(100,85){5}
\Text(100,80)[]{$\tilde{q}_i$}
\DashLine(50,50)(100,50){5}
\Text(100,20)[]{$\tilde{q}_i$}
\DashLine(50,15)(100,15){4}
\Text(70,60)[]{$\Phi$}
\Text(120,50)[]{$+$}
% 2nd Diagram
\Photon(150,85)(200,85){4}{6}
\Text(165,75)[]{$\gamma$}
\Photon(150,15)(200,15){4}{6}
\Text(165,25)[]{$\gamma$}
\DashLine(200,85)(200,15){5}
\DashLine(200,85)(250,85){5}
\Text(250,80)[]{$\tilde{q}_i$}
\DashLine(215,15)(250,50){5}
\Text(250,20)[]{$\tilde{q}_i$}
\DashLine(200,15)(250,15){4}
\Text(240,60)[]{$\Phi$}
\Text(270,50)[]{$+$}
%3d diagram 
\Photon(300,85)(350,85){4}{6}
\Text(315,75)[]{$\gamma$}
\Photon(300,15)(350,15){4}{6}
\Text(315,25)[]{$\gamma$}
\DashLine(350,85)(400,85){5}
\DashLine(365,85)(400,50){5}
\Text(400,80)[]{$\tilde{q}_i$}
\DashLine(350,15)(350,85){5}
\Text(400,20)[]{$\tilde{q}_i$}
\DashLine(350,15)(400,15){4}
\Text(380,60)[]{$\Phi$}
\Text(425,50)[]{$+ {\rm cros.}$}
\end{picture}
\end{center}
%%%%%%%%%%%%%%%%%%%%%%%%%%%%%%%%%%%%%%%%%%%%%%%%%%%
\begin{center}
\begin{picture}(430,100)(0,0)
%4d diagram 
\Text(100,50)[]{$+$}
\Photon(115,75)(150,50){4}{6}
\Text(130,85)[]{$\gamma$}
\Photon(150,50)(115,25){4}{6}
\Text(130,15)[]{$\gamma$}
\DashLine(150,50)(185,75){5}
\Text(170,80)[]{$\tilde{q}_i$}
\DashLine(150,50)(185,25){5}
\Text(170,25)[]{$\tilde{q}_i$}
\DashLine(160,58)(185,40){4}
\Text(185,53)[]{$\Phi$}
% 5th Diagram
\Text(200,50)[]{$+$}
\Photon(215,75)(250,50){4}{6}
\Text(230,85)[]{$\gamma$}
\Photon(250,50)(215,25){4}{6}
\Text(230,15)[]{$\gamma$}
\DashLine(250,50)(285,75){5}
\Text(270,80)[]{$\tilde{q}_i$}
\DashLine(250,50)(285,25){5}
\Text(270,25)[]{$\tilde{q}_i$}
\DashLine(260,42)(285,60){4}
\Text(285,53)[]{$\Phi$}
\Text(350,50)[]{$(\Phi=h, H)$}
\end{picture}
\end{center}
%%%%%%%%%
\vspace*{-5mm}
\begin{figure}[htb]
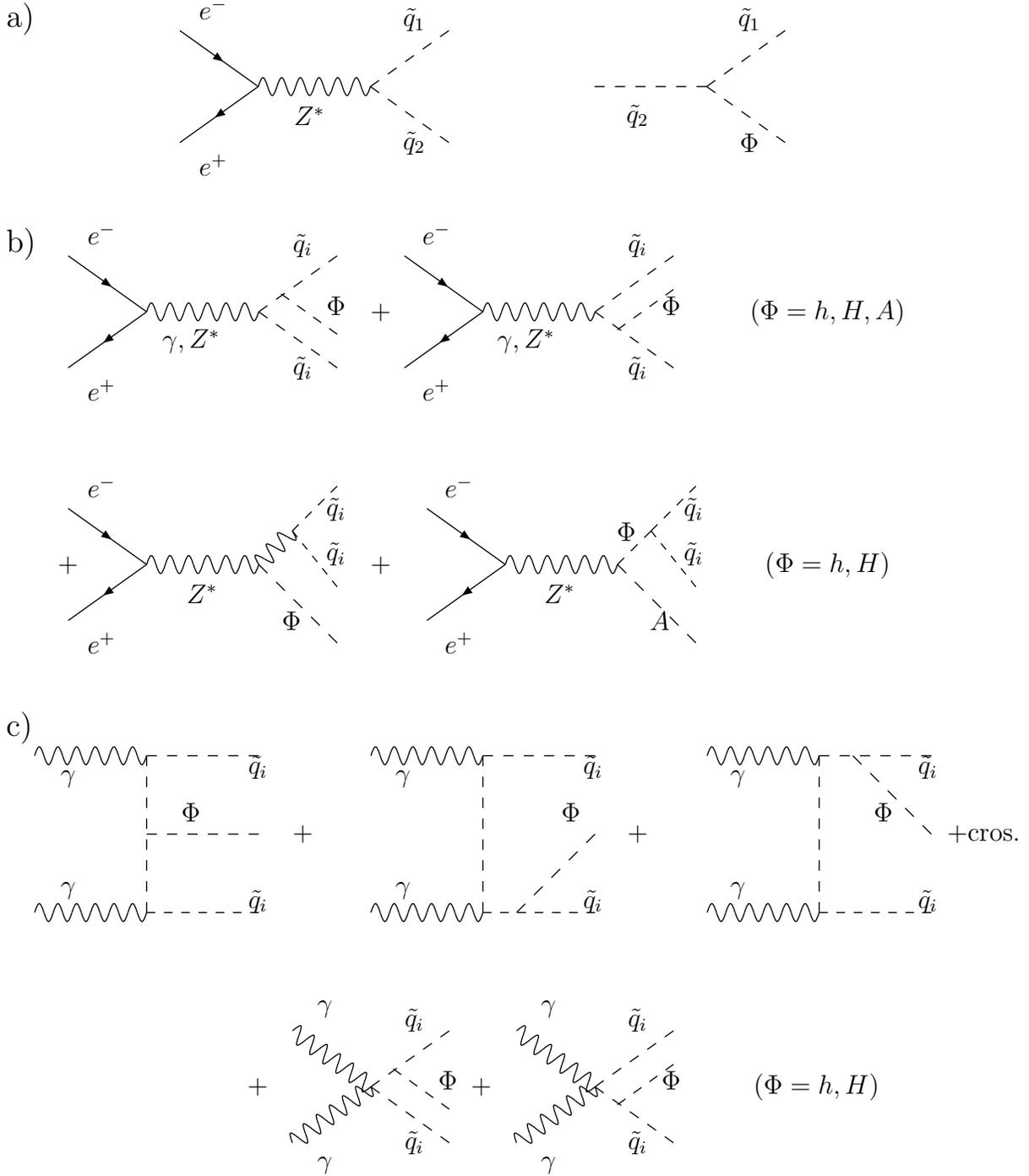

\caption[]{Feynman diagrams for the production of a Higgs
boson $\Phi$ in association with a pair of squarks in $\ee$ and 
$\gamma \gamma$ collisions.}
\end{figure}

\subsection*{4.1 Two--body production and decay} 

If the center of mass energy of the $\ee$ collider is high enough, one can
generate a $\tilde{q}_1 
\tilde{q}_1 h$ final state by first producing a 
mixed pair of squarks, $\ee \ra \tilde{q}_1 \tilde{q}_2$, through the 
exchange of a virtual $Z$--boson, and then let the heaviest squark decay 
into the lightest one and the Higgs boson, $\tilde{q}_2 \ra \tilde{q}_1h$, 
if the splitting between the two squarks is larger than $M_h$;  
Fig.~6a. For $H$ and $A$ bosons final states, a larger splitting in
the stop sector is required. \s

The total cross section for the production of two squarks with different 
masses in $\ee$ annihilation, $\ee \ra \tilde{q}_1 \tilde{q}_2$, including the 
factor of two to take into account the charge conjugated states is given by
\cite{stopd} 
\begin{eqnarray}
\sigma (e^+ e^- \rightarrow \tilde{q}_1 \tilde{q}_2) = \frac{3}{2} \, \sigma_0
\, (a_e^2 +v_e^2) \, a_{12}^2  \, \frac{ \lambda^{3/2}  
(s, m_{\tilde{q}_1}^2, m_{\tilde{q}_2}^2) } {s(s-M_Z^2)^2}
\end{eqnarray}
with $s$ the center of mass energy of the $e^+ e^-$ collider; $\sigma_0 = 
4 \pi \alpha^2/3s$ the QED point cross section and $\lambda$ the usual 
two--body phase--space function, $\lambda (x,y,z) = x^2 +y^2+z^2 -2(xy+xz+yz)$.
The $Z\tilde{q}_1 \tilde{q}_2$ coupling $a_{12}$ is proportional to $\sin 2
\theta_q$, eq.~(15), and in the case where the mixing between squarks is 
absent, this coupling is zero and the cross section vanishes. [The QCD 
corrections to this process, including the SUSY--QCD contributions with
the exchange of gluinos, are available \cite{QCDp} and can be included.] \s

The partial decay width of the heaviest squark into the lightest squark and 
a neutral Higgs boson $\Phi=h,H,A$ is given by \cite{stopd}
\beq
\label{phiborn}
\Gamma (\tilde{q}_2 \ra \tilde{q}_1 \Phi) = \frac{G_F M_Z^4}{2
\sqrt{2} \pi \, m_{\tilde{q}_2}^3 } \, \lambda^{1/2} (m_{\tilde{q}_2}^2,
M_{\Phi}^2, m_{\tilde{q}_1}^2) \; g^2_{\Phi \tilde{q}_1 \tilde{q}_2}
\eeq
where the two--body phase space function $\lambda$ has been given previously. 
[Here also the QCD corrections are available \cite{QCDd} and can be included]. 
As can be seen from inspection of eq.~(\ref{ghtt}), 
 in contrast
with the $g_{\Phi \tilde{q}_1 \tilde{q}_1}$ coupling,  
$\vert g_{\Phi \tilde{q}_1 \tilde{q}_2}\vert $ 
cannot be too large in most of the parameter
space, and in fact this coupling is well dominated by its second term,  
proportional to $\cos 2\theta_q$, except for very large $\tilde A_t$ and/or
when the 
mixing between squarks is maximal,
$\vert\theta_q \vert=\pi/4$, in which case it is anyway 
practically vanishing. \s
 
The branching ratio 
for the $\tilde{q}_2 \ra \tilde{q}_1 \Phi$ decay mode is obtained by dividing 
the partial width by the total $\tilde{q}_2$ decay width. The latter is 
obtained by summing the widths of all possible decay channels. For instance, 
in the case of the more phase--space favored decay channel $\tilde{t}_2 \ra 
\tilde{t}_1 h$, one has to include: decays of the $\tilde{t}_2$ into a bottom 
squark  and a 
charged Higgs or $W$ boson, decays into $\tilde{t}_1$ and a neutral 
Higgs or $Z$ boson, decays into a $b$ quark and charginos, and decays into a 
top quark and neutralinos [or gluinos]. \s

In principle, if phase--space allowed, the cross section for the two--body 
production process times the branching ratio for the two--body decay, should 
be large enough for the final state to be copiously produced. However, as 
discussed above, in a large part of the parameter space 
the cross section times branching ratio will be roughly proportional 
to $\sin 2 \theta_q \times \cos 2 \theta_q$, thus giving small production
rates in the no mixing and maximal mixing scenarii.
In addition, the decay width $\tilde{q}_2 \ra h\tilde{q}_1$ is in general much 
smaller than the $\tilde{q}_2$ decay widths into chargino and neutralinos,
leading to a small branching ratio. Nevertheless, there are regions of 
the MSSM parameter space where the combination $\sin 2 \theta_q 
\times \cos 2 \theta_q$ can be maximal, which occurs typically
for a not too small $m_{\tilde t_L}$--$m_{\tilde t_R}$ splitting and
a moderate $\tilde A_t$, as can be seen from eq.~(\ref{stopmix}). In this case
[which is often realized in particular in the mSUGRA case, as we shall
discuss next] 
the resonant $\tilde t_2$ process may dominate over the non-resonant
$\tilde q \tilde q h$ production and the corresponding rate is visible 
for the high luminosities $\int {\cal L}\dt \sim 500$ 
fb$^{-1}$ expected at  linear colliders \cite{TESLA}. \s

Such a situation is illustrated
 in the case of the $\tilde{t}_1 \tilde{t}_1 h$ final state in 
Fig.~7, where the cross section $\ee \ra \tilde{t}_1 \tilde{t}_2$ times the 
branching ratio BR($\tilde{t}_2 \ra \tilde{t}_1 h)$ is shown as a function of 
the $\tilde{t}_1$ mass at a c.m. energy of $\sqrt{s}=800$ GeV [full lines].
We have chosen a mSUGRA scenario with $\tan\beta=30$, $m_{1/2} =$ 100 GeV,
$A_0 = -600$ GeV and sign$(\mu)= +$. 
[The broken lines show the contribution of the 
non--resonant contributions in this case, which will be discussed in more 
detail later]. 
As can be seen, the cross section can reach the level of 1 fb for relatively
small $m_{\tilde{t}_1}$ values, leading to more than one thousand events 
in the course of a few years, with the expected integrated luminosity of $\int 
{\cal L}\dt \sim 500$ fb$^{-1}$ \cite{TESLA}. Moreover, for the same 
reasons that have
been discussed in previous $pp$ section for the mSUGRA case, 
there are no other constraints e.g.
from CCB or the rho parameter, for the values chosen in our illustration. \s

\begin{figure}[htb]
\vspace*{-.9cm}
\begin{center}
\mbox{
\psfig{figure=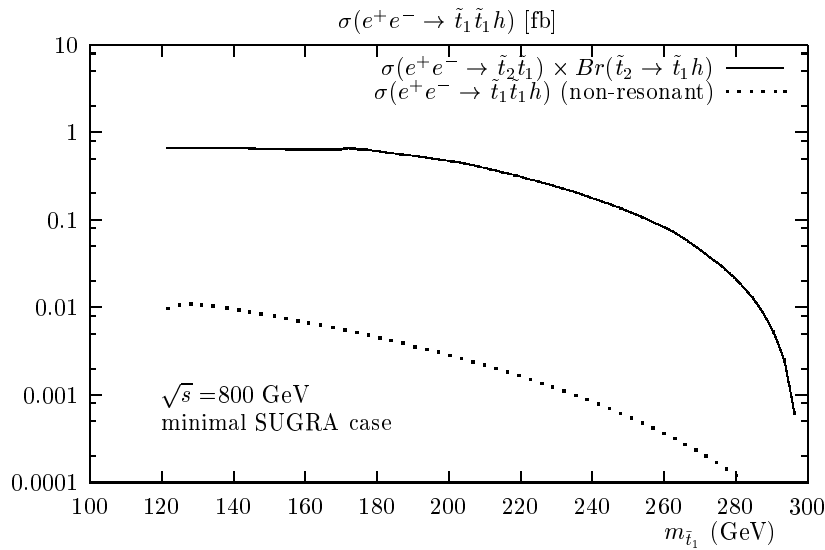,width=15cm}}
\end{center}
\vspace*{-13.cm}
\caption[]{The production cross section $\sigma (e^+e^- \ra \tilde{t}_1 
\tilde{t}_1 h$) [in fb] as a function of $m_{\tilde{t}_1}$ in the mSUGRA 
case; $\tan\beta=$ 30, $m_{1/2} =$ 100 GeV and $A_0 =-600$ GeV.}
\end{figure}

As discussed in the previous section the top squarks in this mass range will 
mainly decay into a charm+neutralino, $\tilde{t}_1 \to c\chi_1^0$, or a $b$ 
quark and a chargino, $\tilde{t}_1 \ra b\chi^+$. In this latter case the 
lightest chargino, $\chi_1^+$  will decay into the LSP and a real or
virtual $W$ boson, leading to the same topology as in the case of the top 
quark decay, but with a large amount of missing energy due to the undetected 
LSP. However, in $\ee$ collisions one can use the dominant decay mode of
the lightest Higgs boson, $h \ra b\bar{b}$. The final state topology will 
consist then of $4b$ quarks [two of them peaking at an 
invariant mass $M_h$, which would be already measured by means of
another process], two [real or virtual] $W$'s and 
missing energy. With the help of efficient micro--vertex detectors, this 
spectacular final state should be rather easy to detect in the clean 
environment of $\ee$ colliders. 
 
\subsection*{4.2 Production in the continuum in e$^+$e$^-$ collisions}

There are three types of Feynman diagrams leading to final states with 
$\tilde{q}_1\tilde{q}_1$ pairs and a CP--even Higgs boson $h$ or $H$,  
in $\ee$ annihilation: Higgs boson emission from the $\tilde{q}_1$ 
states [which are produced through s--channel photon and $Z$--boson 
exchange], Higgs boson emission from the $\tilde{q}_2$ state [which is 
produced with $\tilde{q}_1$ through $Z$--boson exchange], and Higgs boson 
emission from a virtual $Z$ boson which then splits into $\tilde{q}_1 
\tilde{q}_1$ pairs. When the Higgs--$\tilde{q}_1\tilde{q}_1$ 
couplings are large, the two last types of Feynman diagrams give negligible 
contributions since the virtuality of $\tilde{q}_2$ is large and the $Z\tilde{
q}_1\tilde{q}_1$ coupling not enhanced. 
[In the opposite situation where the squark mixing angle $\theta_q$ is non-zero
but moderate, when 
the (resonant) $\tilde q_2$ contribution largely dominates the
cross section, it is then very well approximated by the on--shell 
$\tilde q_1 \tilde q_2$ production with subsequent decay $\tilde q_2 \to 
\tilde q_1 h$, as illustrated in the previous subsection 4.1 
in the mSUGRA case].    
For the CP--odd Higgs boson the first
and last possibilities are absent because of CP--invariance which forbids the
couplings of $A$ to gauge bosons and $\tilde{q}_1 \tilde{q}_1$ pairs. \s

Let us first discuss the process $\ee \ra \tilde{q}_i \tilde{q}_i \Phi$ 
with $\Phi$ the CP--even Higgs boson $h$ or $H$, and $\tilde{q}_i$ any
of the two squarks. In terms of the scaled energies of the two final 
state squarks $x_{1,2}=2E_{1,2}/\sqrt{s}$, the Dalitz plot density for 
the reaction reads 
\beq
\frac{ {\rm d} \sigma } { {\rm d} x_1 x_2} (\ee \ra \tilde{q}_i 
\tilde{q}_i \Phi) = \sigma_0 \frac{\alpha N_c}
{16 \pi s_W^2 c_W^2} \, z \,  \sum _{k= 1,6} A_k 
\eeq
Using the reduced masses $\mu_i= m_{\tilde{q}_i}^2/s , \mu_j= m_{\tilde{q}_j}^2
/s$ [$\tilde{q}_j$ is the virtually exchanged partner of the produced squark 
$\tilde{q}_i$], $\mu_\Phi= M_\Phi^2/s$ and $z=M_Z^2/s$, and the couplings
given in section 2, the various amplitudes squared are as follows: 
\beq
A_1 &=& \left[ e_e^2 e_t^2 + 2\frac{e_e e_t v_e a_{ii} } {1-z} + \frac{ 
(a_e^2+v_e^2) a_{ii}^2 } {(1-z)^2} \right] g^2_{\Phi \tilde{q}_i \tilde{q}_i} 
\left[ \frac{ 2y_1 -1 +4 \mu_i}{y_1^2} - \frac{y_1+y_2 -1+2 x_\Phi}{y_1 y_2} 
\right. \non \\
&& + y_1 \leftrightarrow y_2 \bigg] \non \\
A_2 &=& 
2\left[ \frac{e_e e_t v_e a_{ij}}{1-z} + \frac{(a_e^2+v_e^2)
a_{ii} a_{ij}} {(1-z)^2} \right] g_{\Phi \tilde{q}_i \tilde{q}_j} 
g_{\Phi \tilde{q}_i \tilde{q}_i} \left[ 
\frac{ 2y_1-1 +4 \mu_i}{y_1 (y_1 +\mu_i-\mu_j )} 
- \frac{y_1+y_2 -1 +2 x_\Phi }{y_1 (y_2 +\mu_i -\mu_j) } \right. \non \\ 
&& +  y_1 \leftrightarrow y_2 \bigg] \non \\
A_3 &=& 
\frac{(a_e^2+v_e^2) a_{ij}^2} {(1-z)^2} \, g^2_{\Phi \tilde{q}_i \tilde{q}_j} 
\left[ \frac{ 2y_1-1+ 4 \mu_i}{(y_1+\mu_i-\mu_j)^2} - \frac{y_1+y_2 -1 +2 
x_\Phi }{(y_1+\mu_i - \mu_j) (y_2 +\mu_i -\mu_j) } 
 + y_1 \leftrightarrow y_2 \right] \non  \\
A_4 &=& 
\left[ \frac{e_e e_t v_e a_{ii} g_\Phi }{1-z} + \frac{(a_e^2+v_e^2)
\, a_{ii}^2 \, g^2_\Phi} {(1-z)^2} \right] g_{\Phi \tilde{q}_i \tilde{q}_i} 
\left[\frac{ 4 \mu_i -2 x_\Phi+y_1 -y_2}{y_1 (x_\Phi +2 \mu_i-z)}
+ y_1 \leftrightarrow y_2 \right] \non \\
A_5 &=& 
\frac{(a_e^2+v_e^2) a_{ii} a_{ij}g_\Phi} {(1-z)^2} \, g_{\Phi \tilde{q}_i
\tilde{q}_j} \left[ \frac{ 4 \mu_i -2 x_\Phi+y_1 -y_2}{(y_1+\mu_i - \mu_j) 
(x_\Phi +2 \mu_i-z)} 
+ y_1 \leftrightarrow y_2 \right] \non \\ 
A_6 &=& 
\frac{(a_e^2+v_e^2) a_{ii}^2 g^2_\Phi} {(1-z)^2} 
\frac{ 2 \mu_i - x_\Phi}{(x_\Phi +2 \mu_i-z)^2} 
\eeq
In these equations, $y_{1,2}=1-x_{1,2}$, $x_\Phi=1 - y_1- y_2+ \mu_\Phi 
- 2\mu_i$ and the (reduced) couplings $g_\Phi$ of the $\Phi$ boson to two 
$Z$ bosons [in units of $e M_Z/(s_W c_W)$] 
are given by: $g_h =\sin(\beta-\alpha)$ 
and $g_H = \cos(\beta-\alpha)$. \s

For the final state $\Phi \tilde{q}_1 \tilde{q}_1$ in the case where the $g_{
\Phi \tilde{q}_1 \tilde{q}_1}$ coupling is large [leading to a large splitting
between the $\tilde{q}_1$ and $\tilde{q}_2$ states], the virtuality of the 
squark $\tilde{q}_2$ is large and the $Z\tilde{q}_1\tilde{q}_1$
coupling not enhanced; the Dalitz plot density of the process can be then
approximated by the very simple form 
\beq
\frac{ {\rm d} \sigma } { {\rm d} x_1 x_2} (\ee \ra \tilde{q}_1 \tilde{q}_1 
\Phi) &=& \sigma_0 \frac{\alpha N_c}{16 \pi s_W^2 c_W^2} \, z \left[ e_e^2 
e_t^2 + 2\frac{e_e e_t v_e a_{ii} }{1-z} + \frac{ (a_e^2+v_e^2) a_{ii}^2 } 
{(1-z)^2} \right] g^2_{\Phi \tilde{q}_i\tilde{q}_i} \non \\
& \times & \left[ \frac{ 2y_1 -1 +4 \mu_i}{y_1^2} - \frac{y_1+y_2 -1+2 
x_\Phi}{y_1 y_2} + y_1 \leftrightarrow y_2 \right] 
\eeq

For the associated production of the pseudoscalar Higgs boson with a pair of 
squarks, $\ee \ra \tilde{q}_i \tilde{q}_i A$, the Dalitz plot density has a
much simpler form since only the diagrams with the virtual $\tilde{q}_j$ 
exchange and the one involving the $ZAh(H)$ vertices, are present. It reads:  
\beq
\frac{ {\rm d} \sigma } { {\rm d} x_1 x_2} (\ee \ra \tilde{q}_i 
\tilde{q}_i A) = \sigma_0 \frac{\alpha N_c}
{16 \pi s_W^2 c_W^2} \, z \, \frac{ (a_e^2+v_e^2)}{(1-z)^2} \, 
\sum _{k= 1,3} B_k 
\eeq
where the amplitudes squared, using the same notation as previously, are given 
by
\beq
B_1 &=& 
%\frac{1}{2} 
a_{ij}^2 g^2_{A \tilde{q}_i \tilde{q}_j}\left[ \frac{ 2y_1-1+ 4 \mu_i}{(y_1
+\mu_i-\mu_j)^2} - \frac{y_1 +y_2 -1 +2 x_A }{(y_1+\mu_i - \mu_j) (y_2 +\mu_i 
-\mu_j) }  + y_1 \leftrightarrow y_2 \right] \non  \\
B_2&=& (1-2 y_1- 2y_2 +4 \mu_A) \left[\frac{ g_{ZAh} g_{h \tilde{q}_i 
\tilde{q}_i}}
{x_A +2 \mu_i -\mu_h} + \frac{ g_{ZAH} g_{H \tilde{q}_i \tilde{q}_i} }{x_A +2 \mu_i -\mu_H}  \right]^2 \non \\
B_3 &=& -2 g_{A \tilde{q}_i \tilde{q}_j} 
a_{ij} \left[\frac{ g_{ZAh} g_{h \tilde{q}_i 
\tilde{q}_i}}{x_A +2 \mu_i -\mu_h} + \frac{ g_{ZAH} g_{H \tilde{q}_i \tilde{q}_i}}
{x_A +2 \mu_i -\mu_H}  \right] \non \\
&& \times \left[ \frac{ y_1-2 \mu_A}{(y_1
+\mu_i-\mu_j)} - y_1 \leftrightarrow y_2 \right]
\eeq
with the $Z \Phi A$ couplings 
$g_{ZAh} = \cos (\beta-\alpha)/(2 s_W c_W) $ and $g_{ZAH}=
-\sin (\beta-\alpha)/(2 s_W c_W)$ 
(in units of $e$);
here again $\tilde{q}_j$ is the virtually exchanged partner of the produced 
squark $\tilde{q}_i$ with $\tilde{q}_j \neq \tilde{q}_i$. \s

To obtain the total production cross section $\sigma(\ee \ra \tilde{q}_i 
\tilde{q}_i \Phi)$, one has to integrate over the scaled variables
 $x_{1,2} \equiv 2 E_{1,2}/\sqrt s$ in a way totally similar to the one
described in previous section 3, see eqs.~(\ref{phsp}--\ref{abc}). \s

\begin{figure}[htbp]
\vspace*{-9mm}
\begin{center}
\mbox{
\psfig{figure=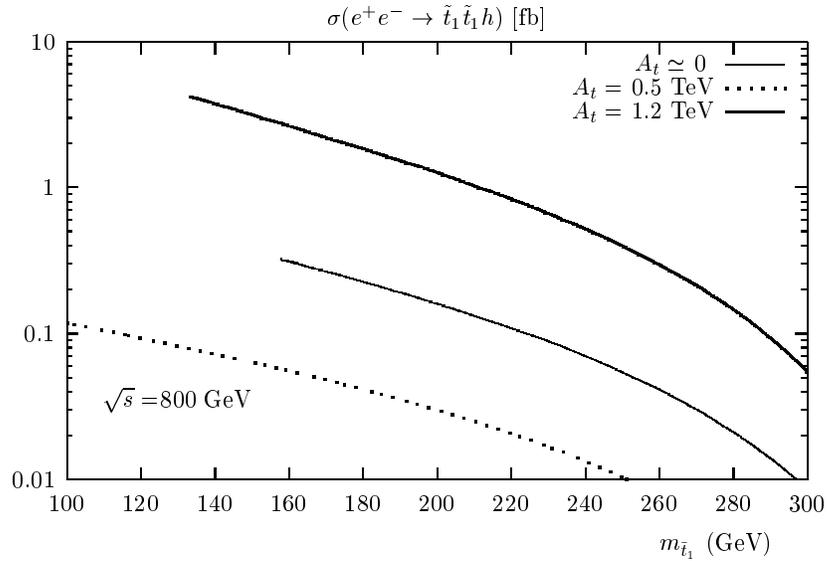,width=15cm}}
\end{center}
\vspace*{-13.5cm}
\caption[]{The cross section $\sigma (e^+e^- \ra \tilde{t}_1 \tilde{t}_1 h$) 
[in fb] as a function of the $\tilde{t}_1$ mass and three different choices
of the other parameters: $\tb=30$, $A_t=$ 0 (0.5) TeV;
$\tb=3$, $A_t=$ 1.2 TeV.} 
\end{figure}
%\vspace*{-1.3cm}

\begin{figure}[htbp]
\vspace*{-.9cm}
\begin{center}
\mbox{\psfig{figure=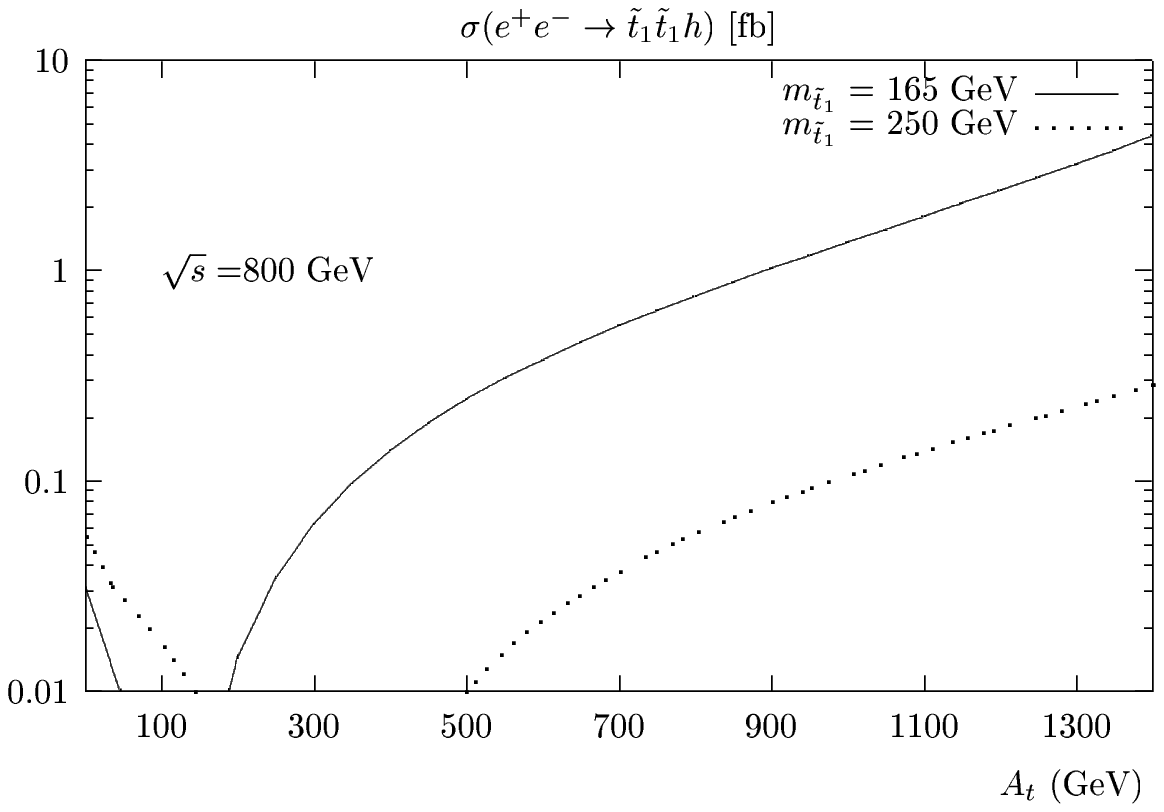,width=15cm}}
\end{center}
\vspace*{-13.5cm}
\caption[]{The cross section $\sigma (e^+e^- \ra \tilde{t}_1 \tilde{t}_1 h$) 
[in fb] as a function of $A_t$ for fixed $m_{\tilde{t}_1}=165$ 
($\tb=3$);
and 250 GeV ($\tb=30)$.}
\end{figure}

To illustrate the magnitude of the cross sections, we show in Figs.~8--9 the 
rates for the $\tilde{t}_1 \tilde{t}_1 h$ final state in the decoupling limit at
$\sqrt{s}=800$ GeV and $\tb=30 (3)$ as a function of the 
$\tilde{t}_1$
mass for $A_t \simeq 0$ 
and $A_t=0.5$ TeV ($A_t =1.2$ TeV) respectively [Fig.~8], 
and as a function of $A_t$ for 
$m_{\tilde{t}_1}=$ 165 GeV ($\tb =3$), and 250 GeV ($\tb =30$) 
[Fig.~9], i.e. for the same scenarii as in 
Figs.~2 and 3 of section 3 [the discussions on the constraints on the 
parameters $A_t$ and $m_{\tilde{t}_1}$ given there apply also in this case]. 
\smallskip

As can be seen the trend is 
similar to the one discussed in the previous section for the associated 
production at proton colliders. For not too large $\tilde{t}_1$ masses 
and large values of the parameter $\tilde A_t$, the production cross sections 
can exceed the value $\sigma( \ee \ra \tilde{t}_1 \tilde{t}_1h) \sim 1$ fb. 
[Note that the cross section for the SM--like process $\ee \ra t 
\bar{t}h$ \cite{eetth} is of the order of 2 fb for $M_h \sim 130$ GeV at a c.m. 
energy $\sqrt{s}=800$ GeV.] 
This provides more than one thousand events in a few years, with a 
luminosity  $\int {\cal L}\dt \sim 500$ fb$^{-1}$: a sample which should be 
sufficient to isolate the final state [with the topologies discussed
in section 4.1] and measure the $g_{\tilde{t}_1 \tilde{t}_1 h}$ coupling with 
some accuracy. \s

At lower energies, the production cross section will be as large as in the
previous case for small $\tilde{t}_1$ masses, but will be limited by phase space
for larger values. This is illustrated in Fig.~10, with the same choice
of parameters as in Fig.~8, but with a c.m. energy of $\sqrt{s}=500$ GeV.  
Thus for very large values of $A_t$, the cross section $\sigma (e^+e^- \ra 
\tilde{t}_1 \tilde{t}_1 h$) can exceed the fb level for stop masses below
$m_{\tilde{t}_1} \lsim 150$ GeV even at a 500 GeV collider. With $\sim 500$
fb$^{-1}$ luminosity, this would hopefully allow to detect the final state
and measure the coupling $g_{h \tilde{t}_1 \tilde{t}_1}$ with some accuracy. 
In the mSUGRA case the cross section generally follows also the same lines as 
in section 
3.2.2 and Fig.~4, i.e. it can be as large as in the case of the unconstrained
MSSM, but in relatively smaller area of the SUSY parameter space, for
reasons that have been given earlier. \s

\begin{figure}[htbp]
\vspace*{-0.9cm} 
\begin{center}
\mbox{
\psfig{figure=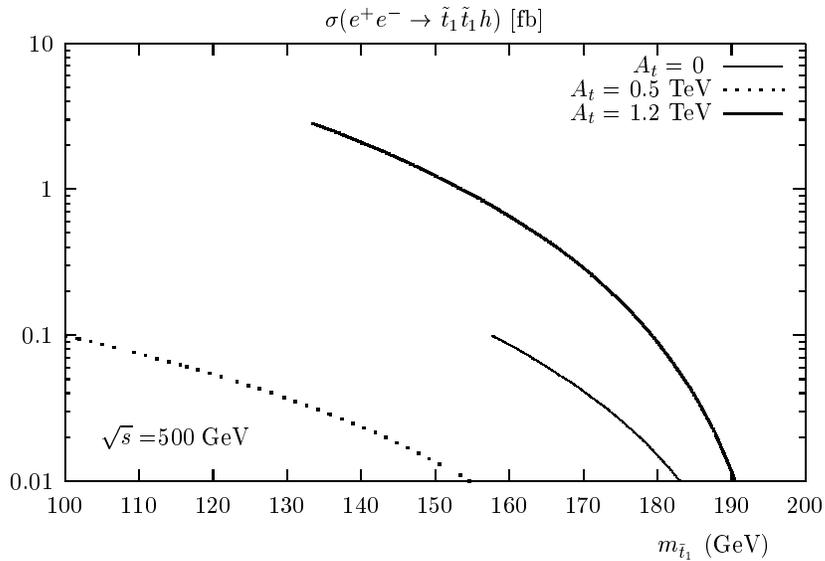,width=15cm}}
\end{center}
\vspace*{-13.cm}
\caption[]{Same as Fig.~8 but with $\sqrt{s}=500$ GeV.}
\end{figure}

Note finally that in the unconstrained MSSM case, the continuum production 
cross section in $\ee$ annihilation $\sigma(\ee \ra \tilde{t}_1 \tilde{t}_1 h$)
is often larger than the resonant cross section for the 
production of $\tilde{t}_1 \tilde{t}_2$ and the subsequent 2--body decay 
$\tilde{t}_2 \ra \tilde{t}_1 h$, but this is not generic. Indeed, 
in a situation where  
both a non-negligible $m_{\tilde t_L}$--$m_{\tilde t_R}$ splitting and
a moderate $\tilde A_t$ occurs, provided there is sufficient phase space
allowed, the production via a resonant $\tilde t_2$ 
becomes competitive and even dominant, as illustrated in a typical
mSUGRA case in the previous subsection 4.1.  

\subsection*{4.3. Production in the continuum in $\gamma \gamma$ collisions}

Future high--energy $\ee$ linear colliders can be turned into high--energy
$\gamma \gamma$ colliders, with the high energy photons coming from Compton 
back--scattering of laser beams. The c.m. energy of the $\gamma \gamma$
collider is expected to be as much as $\sim 80\%$ of the one of the original 
$\ee$ machine. However, the total luminosity is expected to be somewhat 
smaller than the one of the $\ee$ mode, leading to a smaller number of events
for the same cross section.\s

The final state $\tilde{q}_i \tilde{q}_i h$ or $H$ can be generated by 
emitting the $h,H$ bosons from the squark lines in the process $\gamma 
\gamma \ra \tilde{q}_i \tilde{q}_i$; Fig.~6c. The pseudoscalar Higgs boson $A$
cannot be produced in two--photon collisions in association with a pair
of squarks because of CP--invariance; the production of the $A$ boson
is only possible in association with $\tilde{q}_1 \tilde{q}_2$ states
which is less favored by phase space. \s

The differential cross section of the subprocess $\gamma \gamma \ra  
\tilde{q}_i \tilde{q}_i \Phi$ with $\Phi$ a CP--even Higgs boson is given by
\beq
d\sigma (\gamma \gamma \ra \tilde{q}_i \tilde{q}_i \Phi)= \frac{1}{2 s}
\times d{\rm PS} \times |M|^2_{\gamma \gamma} 
\eeq
where dPS is the same element of the 3--body phase space 
discussed in section 3 for
hadron colliders, and $|M|^2_{\gamma \gamma}$ is the amplitude squared of the 
subprocess; taking the same convention as in section 3 for the momenta of the 
initial and final particles, and averaging over spin, the latter is given by
\beq
|M|^2_{\gamma \gamma}= \frac{1}{4} \, e_{q}^4 \, 
(4\sqrt{2} \; G_F \; M_Z^4 \; g^2_{\Phi \tilde{q}_i \tilde{q}_i} \; e^4)
\left[ g_1 + g_2 A_1 +  
g_3 A_2 + g_4 A_1 A_2  +g_5 A_1^2  + g_6 A_2^2 \right]
\eeq
where $A_{1,2}$ have been defined in eq.~(19); in terms of the variables 
$y_{1,2}, x,y$ defined also in eq.~(19), and ${s}, m_{\tilde{q}_i}$ 
the c.m. energy of the $\gamma \gamma$ subprocess and the squark mass
respectively, the coefficients $g_i$ read: 
\beq
g_1 &=&\frac{1}{{s}^2 y_1 y_2} \bigg[2 (p_1.p_2 {s} - 2 x)^2 + 
\frac{(m_{\tilde{q}_i}^2 {s} - 4 y_1) (m_{\tilde{q}_i}^2 {s} 
- 4 y_2) (x^2 - 2 y_1 y_2)}{y_1 y_2} \bigg]
\nonumber \\
g_2 &=&\frac{1}{{s}^2 y_1 y_2^2} \bigg[4 (p_1.p_2 {s} - 2 x) x (-
m_{\tilde{q}_i}^2 p_2.q_{+} {s} + (4 p_2.q_{+} + {s}) y_2) \bigg]
\nonumber\\
g_3 &=& \frac{1}{{s}^2 y_1^2 y_2} \bigg[4 (p_1.p_2 {s} - 2 x) x (
m_{\tilde{q}_i}^2 p_1.q_{+} {s} + (-4 p_1.q_{+} + {s}) y_1) \bigg]
\nonumber \\
g_4  &=& \frac{8}{{s}^2 y_1 y_2} \bigg[ -(p_1.q_{+} p_2.q_{+} 
(p_1.p_2 {s} - 2 x)^2)  \nonumber \\
   &&    + {s} (-2  y_1 y_2 (2 (p_1. q_{+} - p_2. q_{+}) - {s}) - 
m_{\tilde{q}_i}^2 {s} (p_2.q_{+} y_1 - p_1. q_{+} y_2)) \bigg]
\nonumber  \\
g_5 &=& 4 \bigg[2 + \frac{p_2. q_{+} (m_{\tilde{q}_i}^2 {s} 
- 4 y_2) (m^2_{\tilde{q}_i} p_2\cdot q_{+} {s} - 
2  y_2 (2 p_2. q_{+} + {s}) )}{{s}^2 y_2^2} \bigg] \nonumber \\
g_6 &=& 4 \bigg[2 + \frac{p_1.q_{+} (m_{\tilde{q}_i}^2 {s}
 - 4 y_1) (m_{\tilde{q}_i}^2 p_1.q_{+} {s} + 2  y_1 (-2 p_1.q_{+} + 
{s}) )}{{s}^2 y_1^2} \bigg]
\eeq 
Integrating over the phase--space in a similar way as the one 
discussed in section 3, one obtains the 
total production cross section of this subprocess. One can then convolute
 with the 
photon spectra [some examples of spectra can be found in Ref.~\cite{laser}
for instance] to obtain the final cross section $\sigma (\gamma \gamma \ra 
\tilde{q}_i \tilde{q}_i \Phi)$. In the following, we will not use
any photon spectrum for simplicity, we will just exhibit and discuss
the production cross section for the subprocess. \s

The total cross section for the process $\gamma \gamma \ra \tilde{t}_1 
\tilde{t}_1 h$ is shown in Fig.~11 at a two--photon c.m. energy $\sqrt{s}_{
\gamma \gamma}  \lsim 0.8 \sqrt{s}_{ee} =600$ GeV and as a function of the 
$\tilde{t}_1$ mass, without convolution with the photon spectrum and with 
the same inputs and assumptions as in Fig.~8 to compare with the $\ee$ mode. 
Because the c.m. energy of the $\gamma \gamma$ collider is only $\sim 80\%$ 
of the one of the original $\ee$ machine, the process is of course less 
phase--space favored than in the $\ee$ mode. Nevertheless, the cross section for
the $\tilde{t}_1 \tilde{t}_1h$ final state is of the same order as in the 
$\ee$ mode for c.m. energies not too close to the kinematical threshold, and 
the process might be useful to obtain complementary information since it does 
not involve the $Z$--boson and $\tilde{t}_2$ exchanges.  If the luminosities
of the $\gamma \gamma$ and $\ee$ colliders are comparable, a large number
of events might be collected for small stop masses and large $A_t$ values. 

\begin{figure}[htb]
\begin{center}
\mbox{
\psfig{figure=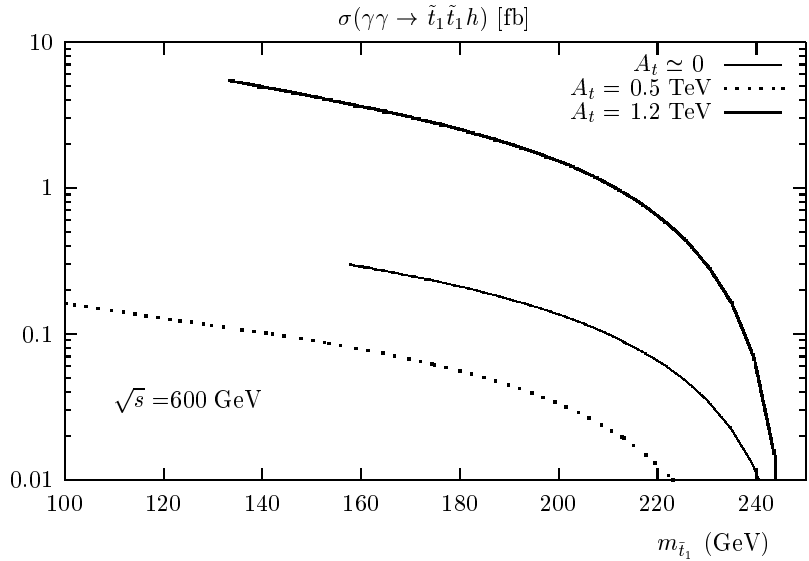,width=15cm}}
\end{center}
\vspace*{-13cm}
\caption[]{The cross section $\sigma (\gamma\gamma \ra \tilde{t}_1 
\tilde{t}_1 h$) in $\gamma \gamma$ collisions [in fb] at a center of mass 
energy $\sqrt{s_{\gamma \gamma}} =600$ GeV as a function of $m_{\tilde{t}_1}$. 
The other parameters have the same values as in Fig.~10.}
\end{figure}

\section*{5. Conclusions} 

We have calculated the cross sections for 
 the production of neutral Higgs particles in association with 
the supersymmetric scalar partners of the third generation quarks, $\ee \ra
\tilde{q}_i \tilde{q}_i \Phi$, at future high--energy hadron colliders and 
at $e^+e^-$ linear machines, in both the $\ee$ annihilation and the $\gamma 
\gamma$ fusion mode. Complete but 
rather simple analytical formulae for the  squared amplitudes of these 
three--body processes are given in the case where the two final state squarks 
have equal mass. In the case of $\ee$ collisions, the production of both the 
CP--even and CP--odd Higgs bosons were calculated, while in $pp$ and $\gamma 
\gamma$ collisions only the production of the CP--even Higgs particles has 
been discussed, since pseudoscalar Higgs bosons cannot be produced at 
tree--level with a pair of equal mass squarks at these colliders. \s

In the framework of the Minimal Supersymmetric extension of the Standard Model, 
we have then investigated the magnitude of the cross sections at the upgraded 
Tevatron and the LHC for hadron machines, and at a future $e^+e^-$ linear 
machine with c.m. energies in the range $\sqrt{s}=$500--800 GeV; in the
latter case, the $\gamma \gamma$ option of the collider has also been 
considered. In this numerical analysis, we focussed on the production of the 
lightest CP--even Higgs boson in the decoupling regime, in association with 
a pair of the lightest top squarks. Final states with the other Higgs bosons 
or with $\tilde{q}_1 \tilde{q}_2$ pairs should be phase--space suppressed. \s

At the LHC,  the production cross sections can be rather substantial, 
especially in the case of rather light top squarks, $m_{\tilde{t}_1}
\lsim 250$ GeV and large trilinear coupling $\tilde A_t \gsim 1$ TeV. 
In this
case the production rates can exceed the one for the standard--like 
production  of the $h$ boson in association with top quarks, $pp \ra t\bar{t}
h$, which is expected to provide a signal of the $h$ boson in the  $\gamma 
\gamma l^\pm$ channel.  Although a  detailed Monte-Carlo analysis, which is  
beyond the scope of this paper, will be required to assess the importance of 
this signal and to optimize the cuts needed not to dilute the contribution of 
the $\tilde{t}_1 \tilde{t}_1h$ final states, it turns out that in a substantial 
area of the MSSM parameter space, the contribution of the top squark to the 
$\gamma \gamma l^\pm$ signal can significantly enhance the potential of the 
LHC to discover the lightest MSSM Higgs boson in this channel. \s

At $\ee$ colliders with c.m. energies around $\sqrt{s}=800$ GeV and with
very high luminosities $\int {\cal L}\dt \sim 500$ fb$^{-1}$, the process
$\ee \ra \tilde{t}_1 \tilde{t}_1h$ can lead to several hundreds of 
events, since the cross sections [in particular for rather light top squarks, 
$m_{\tilde{t}_1} \lsim 250$ GeV and large trilinear coupling, 
$\tilde A_t \gsim 1$ 
TeV] can exceed the level of a 1 fb, and thus the rate for the standard--like
process $\ee \ra t\bar{t}h$. In the case where the top squark decays into 
a $b$ quark and a real/virtual chargino, the final state topology 
[with $4b$ quarks, missing energy and additional jets or leptons]  
will be rather spectacular and should be easy to be seen experimentally,
thanks to the clean environment of these colliders.  
In the $\gamma \gamma$ option of the $\ee$ collider, the cross sections
are similar as previously far from the particle thresholds, but are suppressed 
for larger masses because of the reduced c.m. energies. For $\gamma \gamma$
luminosities of the same order as the original $\ee$ luminosities, the
$\tilde{t}_1 \tilde{t}_1h$ final state should also be observable in the
two--photon mode, at least in some areas of the MSSM parameter space. \s

The production cross section of the $\tilde{t}_1 \tilde{t}_1 h$ final state
is directly proportional to the square of the $\tilde{t}_1 \tilde{t}_1 h$
couplings, the potentially largest electroweak coupling in the MSSM. 
Analyzing these final states at hadron or electron--positron machines will 
therefore allow to measure this coupling, opening thus a window to probe 
directly some of the soft--SUSY breaking scalar potential.  

\bigskip

\subsubsection*{Acknowledgements:} 

We thank M. Drees and M. Spira for discussions, and D. Denegri and E. 
Richter--Was for 
their interest in this problem. This work has been performed in the framework 
of the ``GDR--Supersym\'etrie"; discussions with some members of the GDR
are acknowledged.

\end{document}